\documentclass[a4paper,fleqn,usenatbib]{mnras}

\usepackage{newtxtext,newtxmath}
\usepackage{ulem}
\usepackage{xspace} % for spaces after macros

\usepackage[T1]{fontenc}
\usepackage{ae,aecompl}
\usepackage{xcolor}
\usepackage{graphicx}	% Including figure files
\usepackage{enumitem}
\usepackage{hyperref}
\setlist[enumerate,1]{start=0}

\definecolor{orange}{rgb}{.9,.6,0}
\definecolor{skyblue}{rgb}{.35,.7,.9}
\definecolor{bluishgreen}{rgb}{0,0.6,0.5}
\definecolor{yellow}{rgb}{0.95,0.9,0.25}
\definecolor{blue}{rgb}{0,0.45,0.7}
\definecolor{vermillon}{rgb}{0.8,0.4,0}
\definecolor{reddishpurple}{rgb}{0.8,0.6,0.7}

\newcommand{\tcirc}{\tau_{\rm circ}\xspace}
\newcommand{\tsyn}{ \tau_{\rm sync}\xspace}
\newcommand{\Msun}{{\rm M}_\odot\xspace}
\newcommand{\feh}{[\mathrm{Fe}/\mathrm{H}]\xspace}
\newcommand{\binc}{\textsc{binary\_c}\xspace}
\newcommand{\mint}{\textsc{mint}\xspace}
\newcommand{\mesa}{\textsc{mesa}\xspace}
\newcommand{\bse}{\textsc{bse}\xspace}

\usepackage{xcolor}

\graphicspath{ {./figures/} }

\title[Tides in \binc populations of open clusters]
{Detailed equilibrium and dynamical tides: impact on circularization and synchronization in open clusters}

\author[G.M. Mirouh et al.]{Giovanni M. Mirouh$^{1,2}$\thanks{Email: gmm@ugr.es},
David D. Hendriks$^{1}$,
Sophie Dykes$^{1}$,
Maxwell Moe${^3}$,
\newauthor Robert G. Izzard$^{1}$
\\
$^{1}$Astrophysics Research Group, Faculty of Engineering and Physical Sciences, University of Surrey, Guildford, GU2 7XH, United Kingdom \\
$^{2}$Física Teórica y del Cosmos Department, Universidad de Granada, Granada, Spain\\
$^{3}$Steward Observatory, University of Arizona, 933 N. Cherry Ave., Tucson, AZ 85721, USA}

\date{Accepted 2023 July 4. Received 2023 July 4; in original form 2021 September 29}

\pubyear{2023}

\begin{document}
\label{firstpage}
\pagerange{\pageref{firstpage}--\pageref{lastpage}}
\maketitle

\begin{abstract}
  Binary stars evolve into chemically-peculiar objects and are a major driver
  of the Galactic enrichment of heavy elements. During their evolution they
  undergo interactions, including tides, that circularize orbits and
  synchronize stellar spins, impacting both individual systems and stellar
  populations.  Using Zahn's tidal theory and \mesa main-sequence model grids,
  we derive the governing parameters $\lambda_{lm}$ and $E_2$, and implement
  them in the new \mint library of the stellar population code \binc.  Our
  \mint equilibrium tides are 2 to 5 times more efficient than the ubiquitous
  \bse prescriptions while the radiative-tide efficiency drops sharply with
  increasing age. We also implement precise initial distributions based on
  bias-corrected observations.  We assess the impact of tides and initial
  orbital-parameter distributions on circularization and synchronization in
  eight open clusters, comparing synthetic populations and observations
  through a bootstrapping method.  We find that 
  changing the tidal prescription yields no
  statistically-significant improvement as both calculations typically lie within
  0.5$\sigma$.  The initial distribution, especially the primordial 
  concentration of systems at $\log_{10}(P/{\rm d})
  \approx 0.8, e\approx 0.05$ dominates the statistics even when artificially
  increasing tidal strength. This confirms the inefficiency of tides on the
  main sequence and shows that constraining tidal-efficiency parameters using
  the $e-\log_{10}(P/{\rm d})$ distribution alone is difficult or impossible.
  Orbital synchronization carries a more striking age-dependent signature of
  tidal interactions. In M35 we find twice as many synchronized rotators in our
  \mint calculation as with \bse.  This measure of tidal efficiency is
  verifiable with combined measurements of orbital parameters and stellar
  spins.
\end{abstract}


\section{Introduction}
Multiple systems are commonplace among observed stars: about 35\% of solar-type
stars are in multiple systems, this fraction rising to more than 70\% in O-type
stars \citep{sana12, MS17}.  The presence of a companion can have a significant
impact on the evolution of both stars and is necessary to explain many
astrophysical events and the generation of carbon-enhanced metal- poor (CEMP)
and barium stars, fast rotators, X-ray binaries, novae\ldots \citep{DMI17}.
Tides circularize and shrink orbits while stellar rotation rates synchronize
with the orbit, making them a crucial ingredient of binary evolution.  Studying
stellar populations also offers a way to constrain tides.  Notably, open
clusters are coeval populations of isolated binary systems, numerous
measurements of their orbital parameters make them an interesting laboratory to
assess when and how efficiently tides act.  This work thus focuses on the
derivation of accurate tidal dissipations on the main sequence, which are
expected to modify the orbital parameters of stellar systems, and the study of
both individual binary systems and stellar populations.

Tides in binary systems are divided into two components: the equilibrium tide
and the dynamical tide \citep[we refer the reader to the reviews by][]{zahn08,
ogilvie14}.  The equilibrium tide results from the distortion induced by the
companion's gravitational pull. The resulting bulge rotates with the star
inducing dissipation through friction. This mechanism is efficient in stars
with an outer convective envelope \citep{zahn77, zahn89}.  The dynamical tide
results from the generation of tidally-excited, low-frequency gravity modes of
oscillation at the core-boundary interface.  These oscillations have
periods comparable to that of the orbit.  Resonances thus extract energy from
the orbit that is then dissipated in the stellar envelope through radiative
dissipation or in dissipative shear layers \citep{zahn70, zahn75}. To be
efficient, dynamical tides require a convective core surrounded by a radiative
layer that might in turn be surrounded by an outer convective zone.  Both tidal
mechanisms extract energy from the orbit, resulting in secular changes in the
orbital period $P$, eccentricity $e$ and stellar rotation rates $\Omega$. In
the absence of other interactions, tides typically circularize orbits
($e\rightarrow 0$), while each star tends to spin-orbit pseudo-synchronization
\citep{hut81}.  As systems evolve, close systems circularize first.  In coeval
populations, the period at which no eccentric systems exist -- the cut-off
period -- increases over time \citep[][]{WS02}.  In open clusters in which the
age is determined through turn-off fitting, the cut-off period provides an
observational estimate of the efficiency of tides \citep[][]{MM05}.  Numerous
theoretical formalisms have been developed to explain the observed period
distributions of binary systems in open clusters. Much of this progress
happened over the last decade, reopening a question that is very much in flux.
It is also unclear whether binary stars formed in clusters carry a signature of
their birth conditions.  To test both these aspects, we present here a
derivation of time-dependent tides based on detailed stellar structures that we
implement in the \binc binary evolution code to compute high-resolution
synthetic populations of a variety of open clusters.  We compute tidal
timescales following Zahn's theory \citep{zahn70, zahn75,zahn77,zahn89}.  This
theory introduces a formalism for both equilibrium and dynamical tides,
relating the circularization and synchronization timescales to structure
quantities in both stars, most importantly the coefficients $\lambda_{lm}$ and
$E_2$ whose derivation we summarize in this work. The resulting timescales are
then used in the equations for the secular evolution of orbital parameters
given by \citet{hut81}.

We use the \binc stellar population synthesis code
\citep{binc04,binc06,binc09, binc18} to investigate individual systems and
compute populations.  Since its inception, \binc has been regularly updated to
include new physics such as nucleosynthesis, improved Roche lobe overflow
prescriptions, or rotation \citep[][and references therein]{binc18}.
The rapid evolution algorithm in \binc relies on the ubiquitous \bse parameters
obtained through a series of fits obtained from stellar models \citep{hurley00,
hurley02}. These fitting relations of the stellar mass and age allow for the
rapid evolution of single and binary stars.
In our latest developments of \binc that we call \mint (for
\textit{Multi-object INTerpolation}), we implement a new interpolation approach
based on grids of models over an extensive range of masses and metallicities.
These grids include all the parameters necessary for the main-sequence
evolution, including tides and nucleosynthesis, and are constructed with the
\mesa stellar evolution code \citep{mesa1, mesa2, mesa3, mesa4, mesa5}.  For
each model we calculate the relevant tidal coefficients for both kinds of tides
following the formalism laid out in \citet{zahn77, hut81, zahn89} and
\citet{SIDD13}.  This overhaul of the evolution algorithm will be extended to
later stages of evolution in upcoming papers.  We use stellar populations
obtained with \binc to study both circularization and synchronization
processes.  We investigate eight open clusters that span ages from 4\,Myr to
7\,Gyr and contain a number of binary systems whose orbital parameters have
been measured.  We assess the agreement between our model cluster populations
and corresponding observations through a dedicated bootstrapping method, before
discussing the use of stellar rotation rates and spin-orbit synchronicity as a
possible measure of tidal efficiency.  Throughout this work, we focus on
comparing the \bse and \mint implementations of equilibrium and dynamical tides
to observations.

The paper is structured as follows. We present and justify our prescription for 
tides and detail the differences between the \bse and \mint implementations 
on the evolution of tidal parameters in selected systems in section~\ref{sec:derivation}. In
sections~\ref{sec:populations} and~\ref{sec:poprot} respectively, we
investigate the circularization and synchronization properties of stellar
populations.  We then discuss the implications of our new tidal implementation
in section~\ref{sec:discussion} and summarize our main findings in
section~\ref{sec:conclusions}. Appendices provide 
mathematical details (appendix~\ref{sec:math}) and plots of the computed populations
(appendix~\ref{sec:other_clusters}).

%%%%%%%%%%%%%%%%%%%%%%%%%%%%%%%%%%%%%%%%%%%%%%%%%%%%%%%%%%%%%%%%%%%%%%%%%%%%%%
\section{Derivation of the tidal prescriptions}
\label{sec:derivation}
In this section, we introduce the Zahn formalism of tides that we adopt and
its implications, while technical details are provided in
Appendix~\ref{sec:math}.  To assess the impact of tides on the orbital
evolution of binary stars, we compute the required tidal coefficients from
detailed structures obtained with \mesa. We present these models and give an
overview of the numerical implementation in the \binc population code. 

We also discuss an experiment in which we run a series of systems with different
initial spin and orbital periods to compare the efficiency of \bse and \mint tides for different
initial masses and rotation rates.

\subsection{Our choice of prescriptions: Zahn's formalism}
In this work, we replace the \bse tide prescriptions provided by
\citet{hurley02} with the derivation of \citet{zahn70, zahn75, zahn77, zahn89} and
\citet{hut81}. Despite what the chronology of these works suggests, the \bse
prescriptions are actually a simplification of Zahn's.
Most notably, \bse underestimates tides in close systems
by several orders of magnitude, while their radiative tide implementation is
age-independent and overestimates tidal dissipation as stars evolve on the
main-sequence.

The search for more accurate circularization has led to the development of many
formalisms for both equilibrium and dynamical tides. 
Dynamical tide efficiency is directly related to the rate at which
oscillations dissipate energy in the stellar envelope.  The advent of
asteroseismology has logically ushered an outburst of new calculations for
dynamical tides \citep{willems_etal03, burkhart_etal12}.  Works such as
\citet{T98}, \citet{OL07} or \citet{B20} suggest that damped internal gravity
waves extract energy from the orbit, while others invoke tidally-forced
inertial waves in near-synchronicity systems \citep[e.g.][]{B21}. However, the
timescales upon which tidal forcing takes place are relatively short, and the
coupling itself is quite weak \citep{T98}, unless stellar evolution somewhat
maintains this forcing \citep[through so-called resonance locking,
e.g.][]{SP84,WS02,MF21}. While resonance-locking increases dissipation during
the pre-main-sequence, it is unclear whether it accelerates circularization on
the main sequence significantly \citep{ZW21}.

The equilibrium tide mostly relies on the amount of friction in the stellar
convective envelope. Estimates vary wildly, for instance in the short-period
limit \citep[][a specific case we discuss in
Appendix~\ref{sec:math}]{GN77,VB20}.  \citet{terquem21} and \citet{TM21}
recently suggested that dissipation due to turbulent convection could increase
tidal efficiency, but this idea has been debated since \citep[notably by the
rebuttal of][]{BA21}, while other works emphasize the role of a magnetic field
in increasing dissipation \citep[e.g.][]{wei22}.  A promising study by
\citet{barker22} investigates the impact of inertial wave dissipation in
convective envelopes on equilibrium tides, through calculations similar to
those underlying dynamical tides: their frequency-averaged dissipation rate
seems to yield a good agreement with observations in systems close to
spin-orbit synchronization.

The tension between those different theoretical estimates leads to a
rapidly-changing landscape of tidal theories. However, 
recent works rely on the derivation of the entire oscillation spectrum of
the stars considered. The systematic study of oscillation spectra over the
range of masses and metallicities necessary for this study is a very ambitious
work, even with current computational means, and 
will surely be at the core of highly anticipated future work.
It is worth noting that these formalisms do not yield results
qualitatively different from the formalism we
implement as the conclusions we derive will show \citep{ZW21, TM21}.

As the population synthesis calculations we perform require rapid inferences
over an extended parameter range, we implement Zahn's prescriptions in \mint to
derive circularization and synchronization coefficients owing to their
tractability.  Despite the development of new formalisms, this is the first
implementation of the prescriptions laid out in \citet{zahn89} for population
synthesis. The coefficients thus derived are used in the \binc code in
conjunction with the equations from \citet{hut81} which are necessary to
compute the secular evolution of systems, notably at high eccentricities
\citep[$e>0.3$,][]{TM21}.

\subsection{Our grids of \mesa models}
\label{sec:mesa}
Our derivation of the tidal timescales relies on grids of models of
main-sequence stars constructed using the \mesa stellar evolution code, version
12115.  We make use of the ${\rm d}E/{\rm d}t$ form of the energy equation
paired with gold tolerances, along with both DT2 and ELM equation-of-state
options and type2 opacities (\citealp{mesa5} and references therein).  All our
models rely on a convective mixing length $\alpha_{\rm MLT} = 2$, and
semiconvection is treated following \citet[]{langer85} with $\alpha_{\rm
sc}=0.1$.  We include step overshooting at the convective-core interface
extending from $f=0.05 H_{\rm p}$ inside the convection zone and of thickness
$f_0= 0.33 H_{\rm p}$ with the same diffusion coefficient as convection
\citep[based on the Solar value of][]{jcd11}.
We cover the $0.32 - 100\,\Msun$ mass range at metallicities $Z = 0, 10^{-4},
0.008, 0.012$ and $0.016$, and the extended range $0.1 - 320\,\Msun$ at
$Z=0.02$.  Assuming a reference of $Z=0.02, Y=0.28$ and following the solar
mixture of \citet{GS98}, we include Galactic chemical enrichment using
$dY/dZ=2$ \citep{SB10}.

\begin{figure}
    \centerline{\includegraphics[width=.5\textwidth]{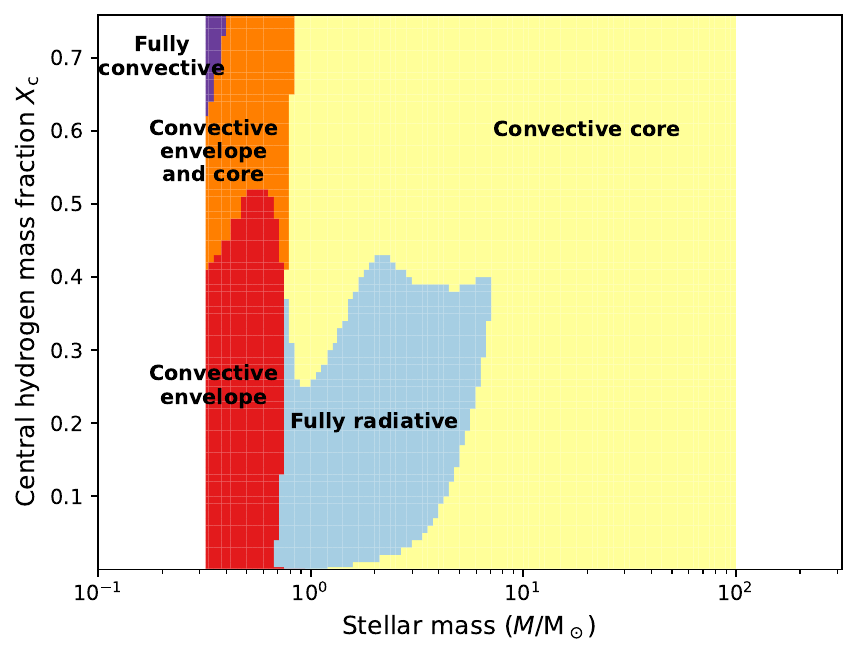}}
    \centerline{\includegraphics[width=.5\textwidth]{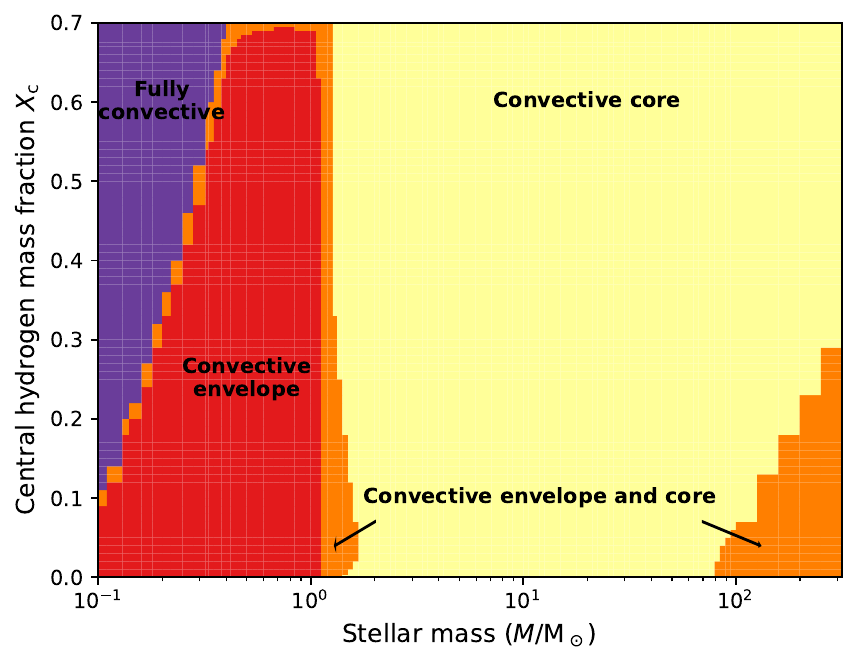}}
    \caption{
    Location of convective regions in stars of metallicities $Z=0$ (top) and
    $0.02$ (bottom), as a function of mass and central hydrogen abundance. The
    ZAMS is at the top of each panel and evolution proceeds vertically
    downwards. Colours denote fully-convective (purple) or fully-radiative
    (blue) stars, or the presence of a convective core (yellow), a convective
    envelope (red) or both a convective core and a convective envelope
    separated by a radiative shell (orange).  Stars with a convective surface
    (red, orange and purple) harbour equilibrium tides while stars with a
    radiative zone around a convective core (orange and yellow) harbour
    dynamical tides.
    }
    \label{fig:radconv}
\end{figure}

Among crucial parameters for tides, the stability of the stellar layers to
convection indicates whether equilibrium or radiative tides dominate.
Fig.~\ref{fig:radconv} shows the distribution of stars featuring a convective
envelope, in which equilibrium tides dissipate energy, and stars with a
convective core in which dynamical tides act.  
At low metallicities, we find stars that are fully radiative on the main
sequence as their convective core disappears.  Zahn's formalism does not
provide a description of tidal dissipation in such stars.  A mechanism that
relies neither on stochastically-excited oscillations nor on main flow viscous
dissipation is needed. \citet{tassoul87, tassoul88} offers such a mechanism
that relies on viscous near-surface boundary-layer dissipation, but its
existence is controversial \citep[][]{rieutord92, RZ97}.  We decide to neglect
it, meaning no tidal dissipation is taken into account in our models of these
fully-radiative stars at low metallicity.  However, we emphasize that none of
the model populations we discuss in this work include such stars.

\subsection{Implementation in \binc}
\label{sec:binc}
Implementing the new tidal prescription in the \binc stellar population synthesis
code \citep{binc04,binc06,binc09, binc18} is part of a larger overhaul of the
code we call \mint.  This change in the algorithm will be the focus of future
papers, but we summarize it here.  To increase the accuracy of the algorithm
that derives stellar parameters used in the code, we replace \bse fitting
relations \citep[][]{hurley00,hurley02} with regularly-spaced grids of \mesa
models that are interpolated linearly. This still allows \binc to rapidly
compute populations as structures are not computed on the fly.  Among the
parameters available in the grids, the coefficients $E$ and $E_2$ yield
$\lambda_{lm}$ and $(k/T)_\mathrm{c}$ following equation~(\ref{eq:lambda_bse}).
Once these coefficients are calculated, they are used in the \citet{hut81} equations 
that govern the evolution of orbital parameters and allow for accurate
calculations of the secular evolution of binary systems for all eccentricities. 
This change is modular, allowing us to swap
easily between \bse and \mint evolution algorithms and tidal prescriptions.
For each population, the computation, output management and data storage is performed with 
the \textsc{binary\_c-python} software package \citep{joss}.

\subsection{Our choice of initial orbital parameters : Moe \& di Stefano (2017)}
\label{sec:initdistro}
In this work, we implement empirical zero-age main-sequence orbital parameter
distributions. Unless otherwise specified, we use a \citet{K01} initial mass
function in conjunction with initial distributions of the mass ratio,
eccentricity and period from \citet{MS17}.  Relying on $\sim30$ observational
surveys based on a variety of techniques, \citet{MS17} performs a careful
correction of observational biases to provide initial binary parameters
distributions. Their study includes stars from the field and from both
solar-like and massive star open clusters to provide tabulated probability
functions of the mass ratio, period and eccentricity.
This empirical distribution arises from the interaction of Kozai-Lidov cycles,
dynamical instabilities and tidal friction during the pre-main-sequence evolution 
\citep{MK18}.

\subsection{Main properties of our new implementation}
\label{sec:secular}
The \mint overhaul of \binc includes changes to both the stellar
evolution algorithm and tides. We find that the changes in the algorithm from
\bse to \mint do not significantly affect the main-sequence evolution of the
stellar structure (e.g. radius and luminosity), but the \mint tides induce
strong differences on the secular orbital parameter evolution.  We assess these
differences through simple experiments we summarize here.

\subsubsection{Efficiency of \mint and \bse tide circularization}
We evolve a set of binary systems with initial eccentricity $e=0.6$ and a range of
initial orbital periods until they exchange mass or
leave the main sequence, whichever comes first.  Evolving these systems
starting at masses $M_1=1\,\Msun$, $M_2=0.5\,\Msun$ and a rotation rate of
$10^{-4}\,\rm{km}\ \rm{s}^{-1}$ for $100\,$Myr, we find that \mint equilibrium
tides circularize all systems with orbital periods shorter than $P \sim 3\,$d while \bse
tides circularize systems with orbital periods shorter than $P = 0.9\,$d.  Over the whole main-sequence
evolution, \mint tides circularize systems up to $P=15\,$d while \bse tides
circularize systems up to $P = 6\,$d.  This comparison shows that \mint
equilibrium tides are more efficient than their \bse counterparts,
circularizing orbits in solar-like binaries more rapidly and affecting
relatively longer-period systems.\\
We repeat the same experiment in systems starting at masses $M_1=50\,\Msun$ and
$M_2=25\,\Msun$, at $e=0.6$ and initial rotation rate of $10^{-4}\,\rm{km}\
\rm{s}^{-1}$.  In this case, \bse dynamical tides circularize systems with
$P<8\,$days over the first Myr while their \mint counterparts circularize
systems with $P<5\,$days. Over the whole main-sequence, circularized systems
reach $P = 25\,$d with \bse tides and $P = 8\,$d with \mint tides. Systems with a
longer orbital period also see their orbit expand near the ZAMS owing to stellar winds.
This experiment confirms that \mint dynamical tides are less efficient than
\bse's. Mathematically, this matches the behaviour of the $E_2$ coefficient:
while on the ZAMS it is similar in both prescriptions, it remains constant in
the \bse calculation but drops significantly in the \mint prescription. This
effect is shown in figure~\ref{fig:E2tracks}.
Age-dependent radiative tides have been used in \citet{yoon2010, SIDD13, qin2018},
we provide a comparison with these calculations in figure~\ref{fig:compare_E2_10}.

\subsubsection{Impact of the initial rotation rate}
\label{sec:rotationrates}
We repeat the above experiment at $M_1=1\,\Msun$, $M_2=0.5\,\Msun$ but vary
the initial rotation rate. We consider four of the \binc possible settings: (i)
a very low rotation rate of $10^{-4}\,\rm{km}\ \rm{s}^{-1}$ that is equivalent
to no rotation, (ii) spin-orbit synchronicity, (iii) breakup, and (iv) with the
\bse mass-dependent initial rotation rate defined as,
\begin{equation}
    v_{\rm rot}(M) = \frac{330 M^{3.3}}{15+ M^{3.45}}\,\rm{km}\ \rm{s}^{-1},
\label{eq:vrot_bse}
\end{equation}
for a given mass $M$ expressed in Solar units \citep[][]{hurley00, lang1992}.
We find no significant impact of the initial rotation on the evolution of
orbital parameters, with circularization happening only slightly faster when
the stars rotate more slowly.

The most notable feature of these tracks concerns systems formed with both
stars at breakup velocity with {$0.2 < \log(P/d) < 0.9$}. They present a
short-lived eccentricity pumping phase on the early main sequence. This can be
traced to equation (10) of \citet{hut81}, in which equilibrium tides provide a
positive contribution to the eccentricity derivative if the stellar angular
frequency exceeds the orbital angular frequency by a factor 5 to 10. 
However, stars undergo magnetic breaking during the pre-main sequence phase and
are not expected to reach the ZAMS at breakup velocities. 
We do not include the pre-main sequence in \mint, but our main-sequence evolution includes magnetic braking 
through the prescription of \citet{andronov03} which is calibrated on open
cluster data and predicts angular momentum loss scaling with $\Omega^3$.

\section{Population synthesis and comparison to cluster observations}
\label{sec:populations}
Binary systems in stellar clusters form with a distribution of initial masses,
eccentricities, and orbital periods.  These stars then evolve through stellar
evolutionary stages while their orbits circularize through tides. As equations
(\ref{eq:tcirc_conv})--(\ref{eq:tsyn_conv}) and
(\ref{eq:tcirc_rad})--(\ref{eq:tsyn_rad}) show, close-period systems
circularize first, so that we observe a dichotomy between close, circular and
wide, eccentric systems. We can define a cut-off period below which all systems
are circular by studying the distribution of binary systems in the
$e-\log_{10}(P/{\rm d})$ plane.  As this cut-off period increases with the
cluster's age, it can be used to infer the age of the population \citep{WS02}.

In this section, we study a sample of open clusters containing binary systems
for which orbital parameters have been measured.  We focus on open clusters
that have a lower stellar density than globular clusters, thus minimizing the
role of N-body interactions.  We compute synthetic populations matching these
clusters with \binc to test initial populations and tidal prescriptions through
their impact on the circularization process.

\subsection{Model populations with \binc}
We compute populations evolving a high number of stars and systems from a given
metallicity and initial orbital-parameter distribution. Each system is evolved
using \binc, relying on either \bse parameters or the interpolation of \mint
grids. We stop the calculation slightly after the documented cluster age and
investigate the eccentricity and orbital period of binary orbits, along with
the stellar rotation rates. The parameter space for these quantities is
divided into bins in which we add the fractional number of stars for each
system at each timestep. In the model populations we present here, we use
950,000 stars for which we track the orbital period, eccentricity, and stellar
spins in units of the critical and pseudo-synchronous rotation rates. 
We store
these quantities in bins of sizes 0.1 for $\log_{10}(P/{\rm d})$ and
$\log_{10}(\Omega/\Omega_{\rm sync})$, and 0.02 for $e$ and $\Omega/\Omega_{\rm
crit}$.  To emphasize the dominant structure of our model populations, we apply
a Gaussian smoothing to the two-dimensional distributions presented in the
figures of this section and the next. This smoothing uses widths 6 and 3 times
the bin sizes on the horizontal and vertical axes, respectively, and is applied
after the statistical calculations we discuss.

\subsection{Goodness-of-fit tests}
\label{sec:stats}
Our model populations provide a distribution of the fractional number of
stars, for instance in the $e-\log_{10}(P/\rm{d})$ plane, that we interpret as a
likelihood map.  To decide whether a set of observations could be drawn from
the synthetic population, we bootstrap two samples from this likelihood map, whose
size matches the number of observed stars for the cluster and period range considered. 
We assess the statistical distance between each of these samples and
observations, and between the two samples, through a two-dimensional
Kolmogorov--Smirnov test \citep[KS test,][]{KS0,KS1}.

This well-established test is a generalization of the one-dimensional KS process
\citep{KS1d} to two dimensions.  The two-sample 1D KS test relies on the
cumulative distribution function of two samples: the statistical distance
between the two samples is defined as the maximum difference between their
cumulative distribution functions, and is directly related to the probability
of the two samples being extracted from a same distribution.

In two dimensions, the key step is to replace the 1D cumulative distribution
function with similar functions computed over the 2D plane by splitting it into
the four natural quadrants around a given point $(x_i, y_i)$,
\begin{equation*}
  (x > x_i, y>y_i), (x <x_i, y>y_i),(x > x_i, y<y_i),(x < x_i, y<y_i).
\end{equation*}
Each quadrant contains part of the samples, yielding cumulative
distribution functions that we compare. The statistical distance is taken as
the largest of the differences between these functions for each of the samples.
\citet{KS1} have shown that this process yields robust inferences when
restricting the choice of $(x_i, y_i)$ to the data points in the samples. From the
statistical distances, it is then possible to retrieve the probability of the
two samples being extracted from the same underlying population through
equations (3), (7), (8) and (9) of \citet{press88}.

In this work, we keep our focus on the statistical distances inferred from
these tests.  First, the statistical distance between the two
bootstrapped samples yields the minimum distance attainable through the KS test.  This
minimum distance follows a Poisson law and serves as a reference value, that we label
as the ``distance to self'' in the rest of this work.  The
same estimator is then used to assess the statistical distance between each of
the two samples and the observed parameters to assess the agreement
between observed and model populations. The distance thus obtained is, by
definition, larger than the Poisson reference.

We repeat this process 1000 times, both for the Poisson reference and the
model--observation statistical distances. These distance estimates distribute
over a Gaussian for which we compute a mean and standard deviation $\sigma$. The
bell-shaped spread of the distances is illustrated, for instance, on 
fig.~\ref{fig:M35stats_nocutoff}, which presents a histogram of the distances in
bins whose width is represented in the top-right corner. The closer
the model--observation distance is to the Poisson reference distance, the more
likely the agreement between the observed and model populations. 

As can be seen in fig.~\ref{fig:m35}, the agreement between observations
and our model populations is driven by two populations: short-period circular
systems and long-period eccentric systems. In order to isolate the
circularization process, we compute the statistical agreement over the whole
population and over a short-period subset, by imposing a cut-off on
$\log_{10}(P/\rm{d})$ that depends on the cluster. It is important to notice
that the common sample size $N$ affects both the distances and their standard
deviations we compute, as they all scale with $\sqrt{N}$: we will thus discuss
the agreement between our populations in units of $\sigma$.

  This approach is fundamentally different from the definition of a cut-off
  period to estimate tidal efficiency, as was done in (e.g.) \citet{MM05}.
  Other studies, such as \citet{Z22} or \citet{bashi23}, extend that cut-off
  period approach by studying the evolution of eccentricity through two
  characteristic periods: one for circular systems and one for more eccentric,
  longer-period systems.  Applying this dichotomic approach to large samples
  (hundreds or thousands of systems) yields crucial statistical insights into
  tidal efficiency.  In this work, we do not perform such separation when
  computing K-S distances.  However, even though the clusters we consider do
  not feature such numbers of binary systems, an exploration of the distinct
  statistics of circular and eccentric systems with our bootstrapping approach
  will be the focus of future work.

\subsection{Our first study case: the cluster M35}
\subsubsection{Impact of the tidal prescription}

\citet{m35_leiner} presents observations of the M35 cluster, a 150 Myr old
cluster with metallicity $\feh = -0.18$. 52 binary systems are detected with
periods 2--4400 days, covering a wide range of eccentricities. Both stars in
each system are on the main sequence, with primary star masses $0.7-1.4\,\Msun$
and no significant information about the mass ratio derived from the
observations \citep[][]{MM05}.  This cluster presents the signature of
circularization processes, with a clear transition from eccentric systems at
periods longer than $\sim 10$ days to only circular orbits at shorter periods.
As such, it is a good test case for our tidal implementation. In this section
we present our population calculations with \binc comparing \mint and \bse
tides.  Starting from the initial parameter distributions described in
section~\ref{sec:initdistro}, we evolve the model populations to an age of
150\,Myr, the age of the cluster documented in the literature.

We study the distribution of stars in the $e-\log_{10}(P/{\rm d})$ plane to
assess the efficiency of circularization and the agreement with observations.

\begin{figure*}
    \centerline{\includegraphics[width=.5\textwidth]{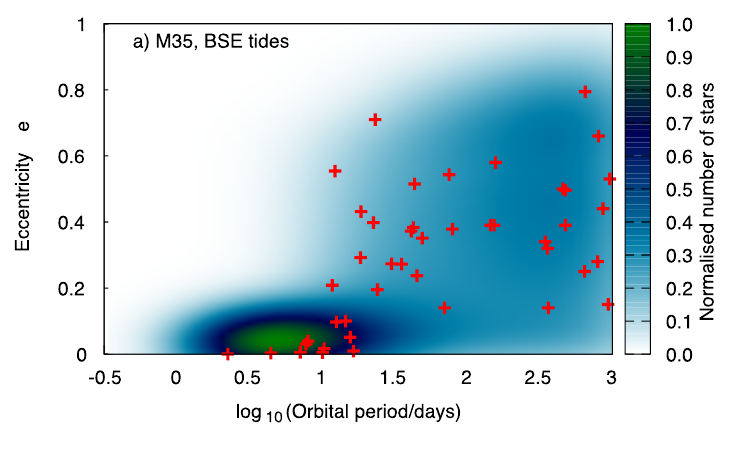}
    \includegraphics[width=.5\textwidth]{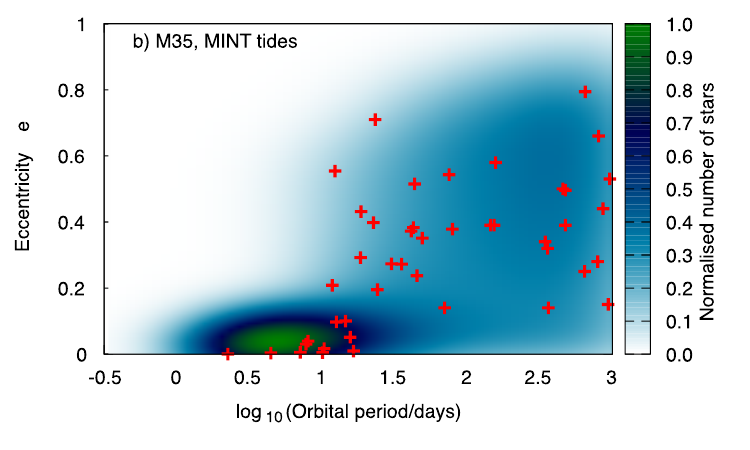}}    
    \caption{
    Comparison between M35 observations (red crosses) and the stellar counts
    calculated populations at 150\,Myr normalized at the highest bin count
    (colour map). Starting with initial distributions from \citet{MS17}, we use 
    tides from \bse (a) and our \mint tides (b).
    }
    \label{fig:m35}
\end{figure*}

Fig.~\ref{fig:m35} shows the $e-\log_{10}(P/{\rm d})$ plane of M35 observations
from \citet{m35_leiner} and our model distributions.  The colour maps indicate
the relative number of our model stars at a given location while the red
crosses are the observed locations of binary systems. Note that the number of
model stars shown in each bin is relative to that of the most populated bin of
either panel.

%%%%%%%%%%%%%%%%%%%%%%%%%%%%%%%%%%%%%%%%%%%%%%%%%%%%%%%%%%%%%%%%%%%%%%%%%%%%%%%%%%%%%%%%%%%%%%%%%%%%%%%

\begin{figure}
    \includegraphics[width=.5\textwidth]{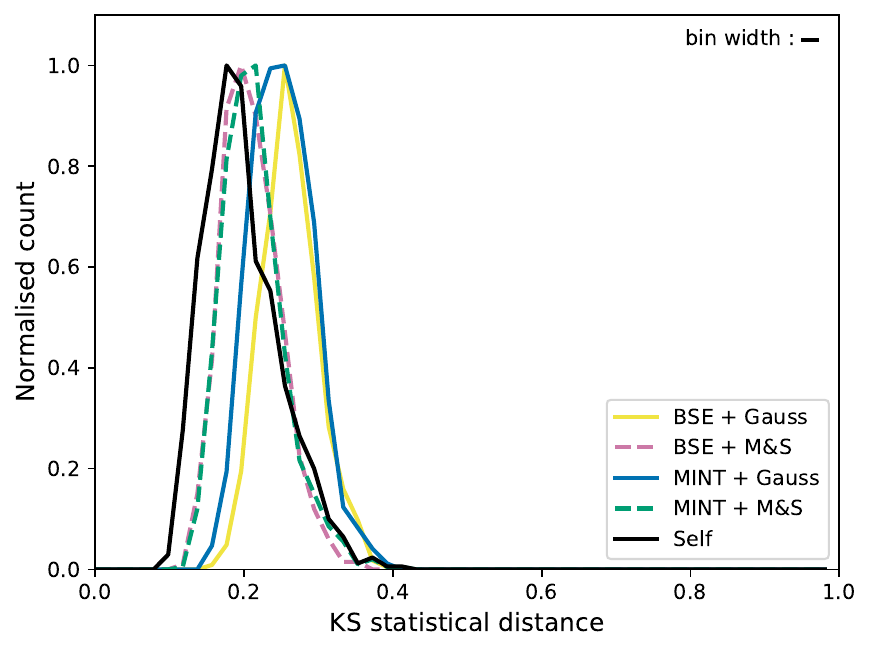}
    \caption{
    Kolmogorov--Smirnov (KS) statistical distance between the whole set of M35
    observations and our corresponding model populations.
    Each coloured line indicates the distance between a model population
    obtained from a physical setup and the observations. 
    Setups include \bse and \mint tides starting from Moe \& di Stefano distributions (M\&S, dashed pink and green resp., see fig.~\ref{fig:m35}),
    and \bse and \mint tides starting from Gaussian distributions (solid yellow and blue resp., see fig.~\ref{fig:m35_gauss}).
    The black curve denotes the reference
    Poisson distance obtained using random samples from one model population.
    The black line in the top-right corner corresponds to the model bin width.
    The statistical mean and standard deviation obtained from these distances
    are reported in Table~\ref{tab:M35stats}.}
    \label{fig:M35stats_nocutoff}
\end{figure}

\begin{figure}
    \includegraphics[width=.5\textwidth]{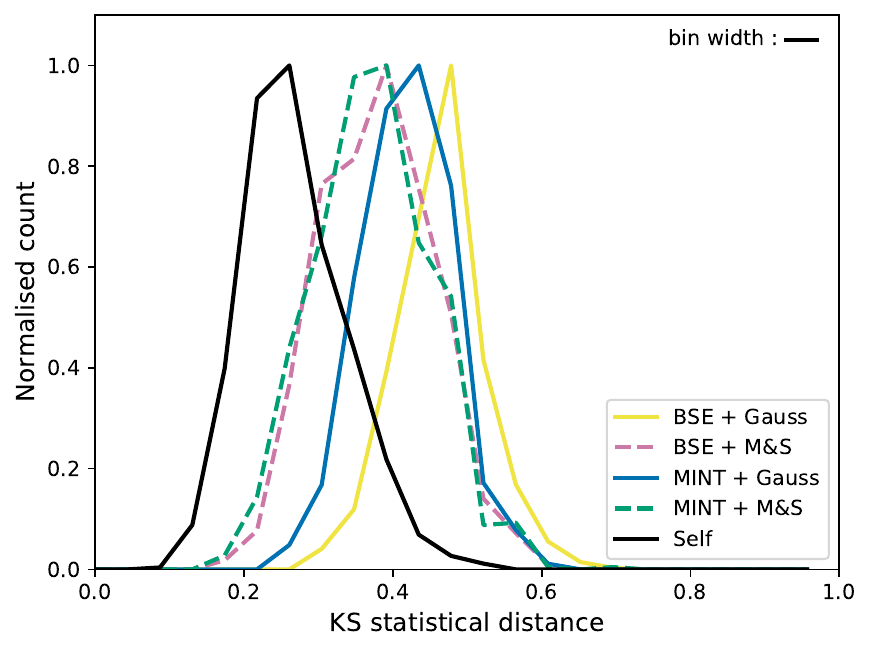}
    \caption{
    As Fig.~\ref{fig:M35stats_nocutoff} restricting the calculation to the
    subset of data with $\log_{10} (P/\mathrm{d}) < 1.7$.  The statistical mean
    and standard deviation obtained from these distances are reported in
    Table~\ref{tab:M35stats}.
    }
    \label{fig:M35_e0_1.7cutoff}
\end{figure}

\begin{table}
\centering
\begin{tabular}{c|c|c|c}
    \hline\\
Cluster and & & & \\
$\log_{10} (P/\mathrm{d})$ range & Tides  & Distance to self  & Distance to obs \\\hline

M35                               & \bse  & $0.208 \pm 0.046$ & $0.220 \pm 0.038$\\
entire sample                      & \mint & $0.209 \pm 0.050$ & $0.224 \pm 0.041$\\ \hline
M35                               & \bse  & $0.290 \pm 0.072$ & $0.396 \pm 0.075$\\ 
$\log_{10} (P/\mathrm{d}) < 1.7$  & \mint & $0.288 \pm 0.066$ & $0.392 \pm 0.078$\\ \hline

\end{tabular}
\caption{Kolmogorov--Smirnov statistical distance estimates for the M35
    model populations starting from Moe \& di Stefano distributions, for \bse or \mint tides,
    for the entire sample or a subset at $\log_{10} (P/\mathrm{d}) < 1.7$.}
\label{tab:M35stats}
\end{table}

\begin{table}
\centering
\begin{tabular}{c|c|c|c}
    \hline\\
Cluster and & & & \\
$\log_{10} (P/\mathrm{d})$ range & Tides  & Distance to self  & Distance to obs \\\hline

M35                               & \bse  & $0.213 \pm 0.047$ & $0.272 \pm 0.037$\\
entire sample                      & \mint & $0.213 \pm 0.044$ & $0.261 \pm 0.040$\\ \hline
M35                               & \bse  & $0.301 \pm 0.069$ & $0.483 \pm 0.064$\\
$\log_{10} (P/\mathrm{d}) < 1.7$  & \mint & $0.296 \pm 0.067$ & $0.436 \pm 0.062$\\ \hline
\end{tabular}
\caption{As Table~\ref{tab:M35stats}, starting from Gaussian initial distributions.}
\label{tab:M35stats_Gauss}
\end{table}

We compare the observations to our two synthetic populations obtained by
changing the tidal prescription.  To describe the circularization process, we
compare observations and model populations through the bootstrapping method
described in section \ref{sec:stats} using both the full set of observations
and a subset of systems with orbital periods shorter than 50 days ($\log_{10}
(P/\mathrm{d}) < 1.7$).  The corresponding distributions of the KS statistical
distance are presented in Figs.~\ref{fig:M35stats_nocutoff}
and~\ref{fig:M35_e0_1.7cutoff} (dashed lines) while their statistical elements
are summed up in Table~\ref{tab:M35stats}.

First, we confirm that the statistical distances obtained by comparing a
computed population to itself do not depend on the underlying physics, but only
on the sample size for each of the runs we have performed.  It serves as a
reference for our other statistical tests.  Using a subset of the observations,
and thus a smaller sample size for the bootstrapping process, generally yields
larger distances and uncertainties but lets us assess the agreement between
populations and observations.

Considering the entire period range, we find a satisfactory agreement between
the observation dataset and the model populations, as the two lie $0.3\sigma$
from the Poisson reference using either \bse or \mint tides.  When focusing on
circularizing systems at $\log_{10} (P/\mathrm{d})<1.7$, the distance rises to
$1.4\sigma$.

In both cases, we find that our model populations are compatible with the
observations. We find no statistically significant difference between the \bse
and \mint prescriptions.  This seems to show that the initial orbital parameter
distribution dominates the circularization distribution on the main sequence.
This is due to the Moe \& di Stefano distribution having a clump of
short-period low-eccentricity systems (centred on $\log_{10} (P/\mathrm{d})
=0.8, e=0.05$) that roughly matches the location of observed circular systems.

%%%%%%%%%%%%%%%%%%%%%%%%%%%%%%%%%%%%%%%%%
\subsubsection{Impact of the initial parameter distributions}
\begin{figure*}
    \centerline{\includegraphics[width=.5\textwidth]{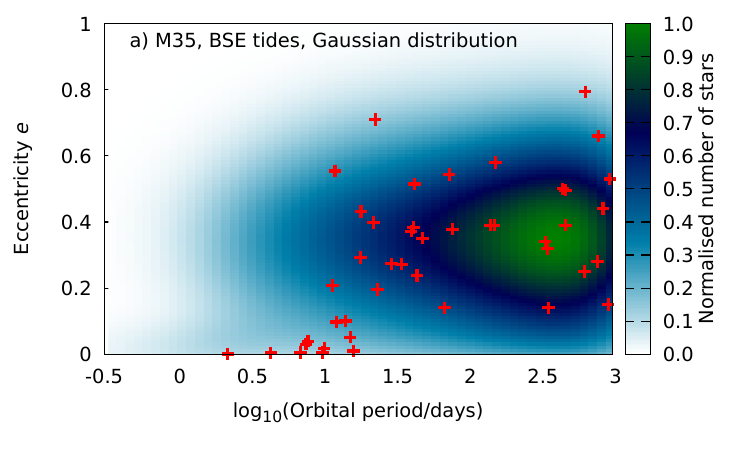}
    \includegraphics[width=.5\textwidth]{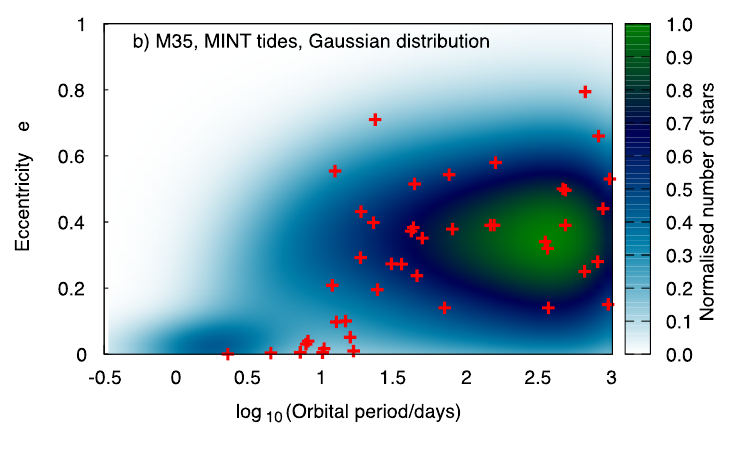}}
    \caption{
    As Fig.~\ref{fig:m35} starting from Gaussian initial distributions.
    }
    \label{fig:m35_gauss}
\end{figure*}

To further assess this last hypothesis, we compute populations starting from a
different, more simple set of initial orbital parameters that do not include a
short-period, low-eccentricity clump.  For this test, we use the initial
parameters suggested by \citet{DM91}: the same Kroupa initial mass function,
along with a flat mass-ratio distribution, a normal distribution of
eccentricities and a log-normal distribution of periods at age zero.  The
Gaussian eccentricity distribution has mean 0.35 and width 0.21, while the
distribution of $\log_{10} (P/\mathrm{d})$ has mean 4.2 and width 4.8.  We
insist that these Gaussian initial orbital period and eccentricity
distributions are not obtained from observations but serve as a proxy for an
initial population without circularized orbits, meant to study the effect of
tides in isolation.  Starting from these distributions, we use \bse and \mint
tides to compute $e-\log_{10}(P/{\rm d})$ distributions to compare with the
observations. These distributions are shown in Fig.~\ref{fig:m35_gauss}.  We
see that \bse equilibrium tides cannot account for the observed
low-eccentricity short-period systems, and while \mint tides yield a small
population of circular close systems, the location and number of stars in this
subset of the parameter space do not match the observed systems.

Gaussian initial distributions deteriorate the agreement between the observed
and model populations significantly, as the statistical elements presented in
Figs.~\ref{fig:M35stats_nocutoff} and~\ref{fig:M35_e0_1.7cutoff} (solid lines)
and Table~\ref{tab:M35stats_Gauss} show.  For the entire sample, the
statistical distance increases from $0.3\sigma$ to $1.4\sigma$ with \bse tides
and to $1.1\sigma$ with \mint tides.  When focussing on the short-period
systems at $\log_{10} (P/\mathrm{d})<1.7$, we find that the distance between
observations and models increases to $2.4-2.8\sigma$.

These distances confirm the significant impact of the \citet{MS17} initial
distributions in improving the agreement between observed and modeled
eccentricities and periods for open clusters, notably thanks to the primordial
population of circular close systems. We find similar results for all the
clusters presented in section~\ref{sec:all_clusters}, but will not discuss them
further owing to the unrealistic nature of the underlying Gaussian
distributions.

%%%%%%%%%%%%%%%%%%%%%%%%%%%%%%%%%%%%%%%%%
\subsection{Other clusters}
\label{sec:all_clusters}
\begin{figure*}
    \includegraphics[width=.75\textwidth]{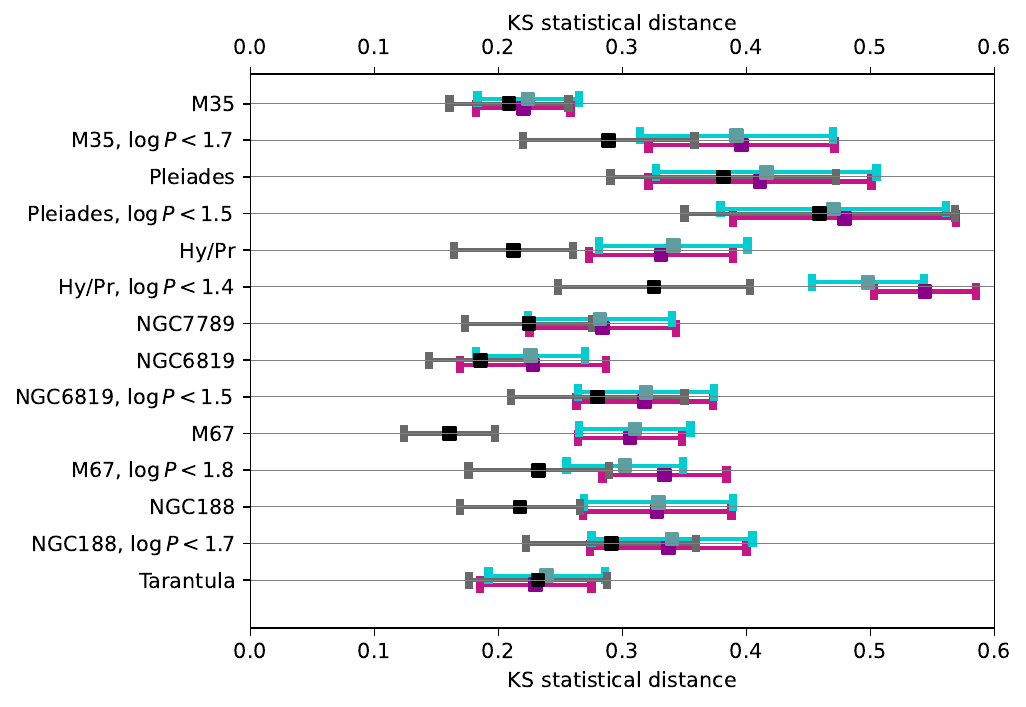}
    \caption{
      Mean and standard deviation for the KS distance estimates between model and observations
      for the samples indicated.
      We plot the distance to self (black), and the distance between the observations and populations
      computed using \bse tides (purple) and \mint tides (blue).
    }
    \label{fig:stats_summary}
\end{figure*}

After having established the method on M35, we apply it to seven other clusters
for which binary populations have been observed to assess whether our updated
initial orbital parameter distributions and tide prescriptions can match
observations.  We list these clusters and their key properties in
Table~\ref{tab:cluster_settings}.  All but one of these clusters contain
main-sequence late-type stars that we present in order of increasing age from
100\,Myr to 7\,Gyr. It is worth noting that while we use six clusters of
late-type main-sequence stars, in which equilibrium tides are expected to
dominate the circularization process, the range of masses, ages, and
metallicities covered lead to a variety of internal structures and tidal
coefficients. The notable exception is Tarantula, a very young cluster of
O-type stars that allows us to assess dynamical tides in massive stars.  In
this section, we discuss the population parameters and their agreement with
observations, $e-\log_{10}(P/{\rm d})$ diagrams and are presented in
appendix~\ref{sec:other_clusters}. The statistical elements obtained for all
clusters are listed in Table~\ref{tab:stats_summary} and shown in
Fig.~\ref{fig:stats_summary}.

\begin{table*}
\centering
\begin{tabular}{c|c|c|c|c|c}
    \hline\\
Cluster            & MS systems           & Age (Gyr)   & Mass range            & $\feh$ & References for observations \\\hline
M35                & 52                   & $0.15$      & $0.7 - 1.4\,\Msun$    & $-0.18$ & \citet{MM05, m35_leiner}            \\
Pleiades           & 13                   & $0.1$       & $0.9 - 1.4 \Msun$     & $+0.042 $ & \citet{Pl_M92, Pl_M97}            \\
Hyades/Praesepe    & 53                   & $0.63$      & $0.5 - 1.5\,\Msun $   & $+0.14,+0.21 $ & \citet{HyPr_G78, HyPr_G81, HyPr_G82, HyPr_G85} \\
                   &                                                     &    &&& \citet{HyPr_M90, HyPr_M92, HyPr_M99}             \\
NGC 7789           & 43                   & $1.6$       & $1.4 - 1.8\,\Msun $   & $+0.02 $ & \citet{ngc7789_N20}\\
NGC 6819           & 68                   & $2.5$       & $1.1 - 1.6\,\Msun $   & $+0.09$ & \citet{ngc6819_M14, ngc6819_h09}                             \\
M67                & 94                   & $4$         & $0.7 - 1.3 \Msun $    & $+0.05 - +0.1$& \citet{m67_G21}\\
NGC 188            & 49                   & $7$         & $0.9 - 1.14 \Msun $  & 0 & \citet{ngc188_G09, ngc188_G12}\\
Tarantula          & 38                   & $\sim 0.004$       & $20 - 80\,\Msun $  & $-0.37$ & \citet{tar_almeida17}                             \\
        \hline
\end{tabular}
\caption{Summary of the cluster observational information used for population synthesis.}
\label{tab:cluster_settings}
\end{table*}

% STATS TABLE WITH ALL STATS TOGETHER
\begin{table*}
\centering
\begin{tabular}{c|c|c|c|c|c}
    \hline\\
Cluster and $\log_{10} (P/\mathrm{d})$ range & Tides  & Distance to self  & Distance to obs \\\hline
Pleiades                      & \bse  & $0.383 \pm 0.092$ & $0.411 \pm 0.090$\\ 
entire sample                 & \mint & $0.380 \pm 0.090$ & $0.416 \pm 0.089$\\ \hline
Pleiades                      & \bse & $0.461 \pm 0.109$ & $0.479 \pm 0.090$\\ 
$\log_{10}(P/{\rm d}) < 1.5$  & \mint & $0.457 \pm 0.109$ & $0.470 \pm 0.091$\\ \hline

Hyades/Praesepe               & \bse    & $0.212 \pm 0.048$ & $0.331 \pm 0.058$ \\
entire sample                 & \mint   & $0.212 \pm 0.048$ & $0.341 \pm 0.060$ \\ \hline
Hyades/Praesepe               & \bse    & $0.330 \pm 0.078$ & $0.544 \pm 0.041$ \\
$\log_{10}(P/{\rm d}) < 1.4$  & \mint   & $0.321 \pm 0.077$ & $0.498 \pm 0.045$ \\ \hline

NGC 7789                      & \bse & $0.223 \pm 0.052$ & $0.284 \pm 0.059$\\
entire sample                 & \mint & $0.226 \pm 0.051$ & $0.282 \pm 0.058$\\ \hline

NGC 6819                      & \bse  & $0.186 \pm 0.042$ & $0.228 \pm 0.044$\\
entire sample                 & \mint & $0.185 \pm 0.041$ & $0.226 \pm 0.044$\\ \hline
NGC 6819                      & \bse & $0.282 \pm 0.071$ & $0.318 \pm 0.055$\\
$\log_{10}(P/{\rm d}) < 1.5$  & \mint & $0.278 \pm 0.069$ & $0.319 \pm 0.055$\\ \hline

M67                           & \bse    & $0.160 \pm 0.037$  & $0.306 \pm 0.042$\\
entire sample                 & \mint   & $0.161 \pm 0.036$  & $0.310 \pm 0.045$\\ \hline
M67                           & \bse & $0.236 \pm 0.057$ & $0.334 \pm 0.050$\\
$\log_{10}(P/{\rm d}) < 1.8$  & \mint & $0.229 \pm 0.056$ & $0.302 \pm 0.047$\\ \hline

NGC 188                       & \bse  & $0.218 \pm 0.049$ & $0.328 \pm 0.060$\\
entire sample                 & \mint & $0.217 \pm 0.048$ & $0.329 \pm 0.060$\\ \hline
NGC 188                       & \bse  & $0.292 \pm 0.068$ & $0.337 \pm 0.063$\\
$\log_{10}(P/{\rm d}) < 1.7$  & \mint & $0.290 \pm 0.069$ & $0.340 \pm 0.065$\\ \hline

Tarantula                     & \bse & $0.234 \pm 0.056$ & $0.230 \pm 0.045$\\
entire sample                 & \mint & $0.230 \pm 0.055$ & $0.239 \pm 0.047$\\ \hline
\end{tabular}
\caption{Mean and standard deviation for the KS distance estimates between model and observations 
  for all the samples considered here. All these calculations rely on Moe \& di Stefano initial
  orbital parameters distributions.}
\label{tab:stats_summary}
\end{table*}

%%%%%%%%%%%%%%%%%%%%%%%%%%%%%%%%%%%%%%%%%%%%%%%%%%%%%%%%%%%%%%%%%%%%%%%%%%%%%%%%%%%%%%%%%%%%%%
\subsubsection{Pleiades}
The Pleiades is a young, 100\,Myr old stellar cluster for which observations by
\citet{Pl_M92, Pl_M97} provide the orbital parameters of 13 binary systems with
masses from $0.9$ to $1.4\,\Msun$.  It has $\feh=+0.042$ and we use $Z=0.016$
for our model population.  We compute model populations for this cluster using
the same approach as for M35 and present the associated period-eccentricity
distributions in Fig.~\ref{fig:pleiades}.  We compute the agreement between the
computed population and the observations, following the bootstrapping method
described above, for both the whole dataset, and a subsample at $\log_{10}
(P/\mathrm{d})<1.5$. 

Over the entire dataset, we find that both populations lie at a distance of
about $0.4\sigma$ from observations, with \bse and \mint tides lying at
$0.1\sigma$ of each other.  The short-period subsample we consider contains 8
systems that are expected to be circularized and have a similar behaviour, as
the population computed with \bse tides lies at $1.8\sigma$, and \mint tides
lower this distance to $1.3 \sigma$.

This shows that the observations of the young Pleiades model populations bear
the signature of the $\log_{10} (P/\mathrm{d})=0.8, e=0.05$ clump in the
initial distribution and have not been impacted by tides in a way that allows
us to significantly assess the best tidal prescription from circularization.
It is also crucial to note that the large standard deviations and Poisson
distances in both calculations  prevent a definite identification of the best
candidate model population when relying on such small sample sizes. 

%%%%%%%%%%%%%%%%%%%%%%%%%%%%%%%%%%%%%%%%%%%%%%%%%%%%%%%%%%%%%%%%%%%%%%%%%%%%%%%%%%%%%%%%%%%%%%%%%%%%%%%%%%%%%%%%%%%%
\subsubsection{Hyades/Praesepe}
Hyades and Praesepe are twin super-solar clusters (${\feh=+0.014}$ and
$+0.021$, respectively) that formed together about $630$\,Myr ago.
Observations from a series of articles referenced in
Table~\ref{tab:cluster_settings} provide the orbital parameters of 53 systems
with masses $0.5-1.5\,\Msun$.  Our model population using $Z=0.02$ is presented
in Fig.~\ref{fig:hypr} This cluster is older than M35 or Pleiades, leaving more
time for tides to act on close systems.

For both the \mint and \bse tides, the model populations we compute lie
$2.1\sigma$ from the Poisson reference.  For a subset of circularizing systems
with $\log_{10} (P/\mathrm{d}) < 1.4$, neither of our model populations match
the observed parameters with the best model lying $3.7\sigma$ away from the
reference. 

This mismatch between the observed systems and our computed populations is due
to the intermediate-period eccentric systems (at $0.7 < \log_{10}(P/d) < 1.2,
e>0.2$) that our calculations do not predict. These peculiar systems were
already highlighted by \citet{DM91}, and impact negatively the calculation of
the circularization period by \citet{MM05}.  They may be explained by the
presence of an outer tertiary companion. Either through Kozai-Lidov
interactions pumping the eccentricity of the inner binary \citep{raghavan10} or
through the interaction of these Kozai-Lidov cycles with tides shrinking the
orbit of originally wider systems \citep{MK18}, triple-star effects lead to
intermediate-period eccentric systems that cannot be explained by binary
evolution alone.

%%%%%%%%%%%%%%%%%%%%%%%%%%%%%%%%%%%%%%%%%%%%%%%%%%%%%%%%%%%%%%%%%%%%%%%%%%%%%%%%%%%%%%%%%%%%%%%%%%%%%%%%%%%%%%%%%%%%%
\subsubsection{NGC 7789}
NGC 7789, presented in \citet{ngc7789_N20}, is a 1.6\,Gyr cluster with $\feh =
+0.02$ in which 43 main-sequence stellar systems are identified in the
$1.4-1.8\,\Msun$ range (Nine, private communication). We compute a model
population at $Z=0.016$ for masses covering this range. 

The distribution of our model population in the $e-\log_{10}(P/{\rm d})$ plane
is shown in Fig.~\ref{fig:ngc7789}. This cluster contains a population of
systems at ${e<0.2}, {\log_{10} (P/\mathrm{d})<1.2}$ surrounding the location
of the clump from Moe \& di Stefano's initial distributions. This concentration
could be attributed to tidal circularization on the main-sequence, but the
statistical distance between observations and model populations is $1.7\sigma$
for both tidal prescriptions, thus showing that the agreement between the
computed population and the observed parameters mostly depends on the initial
conditions, as is the case for the clusters discussed previously. 

%%%%%%%%%%%%%%%%%%%%%%%%%%%%%%%%%%%%%%%%%%%%%%%%%%%%%%%%%%%%%%%%%%%%%%%%%%%%%%%%%%%%%%%%%%%%%%%%%%%%%%%%%%%%%%%%%%%%
\subsubsection{NGC 6819}
NGC 6819 \citep{ngc6819_h09} is slightly metal-rich with $\feh = +0.09\pm0.03$
\citep{ngc6819_b01} and age 2.4\,Gyr.  We resample the 68 main-sequence stars
of \citet{ngc6819_M14} by applying cut-offs $V>14.85$ and $0.7 < (V-I) < 0.95$
to their photometric data.  The systems cover the period range $0.1 <
{\log_{10} (P/\mathrm{d})< 3.6}$ with primary masses $1.1-1.6 \Msun$. We
compute a model population at $Z=0.0175$ for this range of primary masses.

Our model populations in the $e-\log_{10}(P/{\rm d})$ plane are shown in
Fig.~\ref{fig:ngc6819}.  This cluster contains a population of near-circular
systems at ${e<0.1}, {\log_{10} (P/\mathrm{d})<1.2}$ matching Moe \& di
Stefano's initial distributions.  Comparing the whole set of observations to
our computed populations, we find that the populations lie at $1\sigma$ from
each other, with both \bse tides and \mint tides.  Focusing the statistical
inference on the circularizing systems with $\log_{10} (P/\mathrm{d})<1.5$
improves the agreement further, as the populations lie $0.6\sigma$ away from
the observations.  This confirms that the observed distribution of binary
systems can be reproduced when choosing accurate initial distributions, and
that the choice of tidal prescription only has a marginal impact.

%%%%%%%%%%%%%%%%%%%%%%%%%%%%%%%%%%%%%%%%%%%%%%%%%%%%%%%%%%%%%%%%%%%%%%%%%%%%%%%%%%%%%%%%%%%%%%%%%%%%%%%%%%%%%%%%%%%%
\subsubsection{M67}
M67, also known as NGC 2682, is a 4\,Gyr cluster with $\feh$ between $+0.05$
and $+0.1$ in which 94 main-sequence binary systems are observed
\citep{m67_G21}.  These stars are divided between circular systems with
${e<0.05}, {\log_{10} (P/\mathrm{d})<1.2}$ and eccentric systems with $e<0.9,
{\log_{10} (P/\mathrm{d}) > 0.8}$.  They belong to the $0.7-1.3\,\Msun$ range,
which we use for our population study with metallicity $Z=0.0175$.

The model populations we compute are shown in Fig.~\ref{fig:M67}.  We find that
the agreement between our model populations and the 94 observed binary systems
is not satisfactory, as it goes from $3.7\sigma$ with \bse tides to $3.3\sigma$
then using \mint tides.  The relatively poor statistical agreement between our
calculated populations and the observed binary systems of M67 can be attributed
to the 6 long-period eccentric systems ($e>0.7, {\log_{10} (P/\mathrm{d}) >
1.8}$), whose distribution is only marginally matched in our calculations.

Selecting systems with ${\log_{10} (P/\mathrm{d}) < 1.8}$ confirms that the
best statistical agreement is obtained using \mint tides, with model
populations and observations lying $1.55\sigma$ apart.

%%%%%%%%%%%%%%%%%%%%%%%%%%%%%%%%%%%%%%%%%%%%%%%%%%%%%%%%%%%%%%%%%%%%%%%%%%%%%%%%%%%%%%%%%%%%%%%%%%%%%%%%%%%%%%%%%%%%
\subsubsection{NGC 188}
The oldest cluster we consider is NGC 188, at an age of 7\,Gyr and solar
metallicity \citep{ngc188_M04}.  Starting from the photometry of
\citet{ngc188_G09}, we select the main-sequence stars with $V > 15$ and $0.65 <
(B-V) < 0.9$ (Mathieu, private communication).  This leaves us with a sample of
49 stars in the $0.9-1.14$ mass range \citep{ngc188_G09,ngc188_G12}, that we
use for our population along with solar metallicity $Z=0.0142$.

We present our model populations in Fig.~\ref{fig:ngc188}.  We find that the
distance between all 49 observed main-sequence systems and model populations is
of $2\sigma$ with both tidal prescriptions.  When focusing on a subset of close
systems with $\log_{10} (P/\mathrm{d}) < 1.7$, the distance drops to
$0.7-0.8\sigma$.  Despite 7 Gyr of main-sequence evolution, this cluster
carries a strong signature of the initial orbital parameter distribution that
tides cannot dissipate. 

%%%%%%%%%%%%%%%%%%%%%%%%%%%%%%%%%%%%%%%%%%%%%%%%%%%%%%%%%%%%%%%%%%%%%%%%%%%%%%%%%%%%%%%%%%%%%%%%%%%%%%%%%%%%%%%%%%%%
\subsubsection{Tarantula}
\begin{figure*}
    \centerline{\includegraphics[width=.5\textwidth]{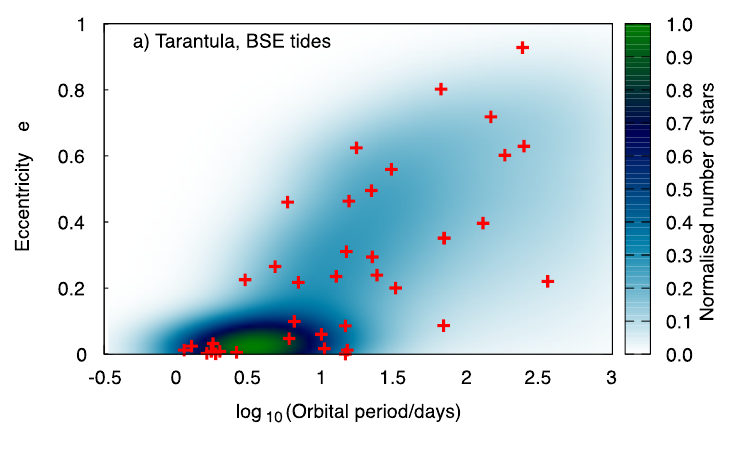}
    \includegraphics[width=.5\textwidth]{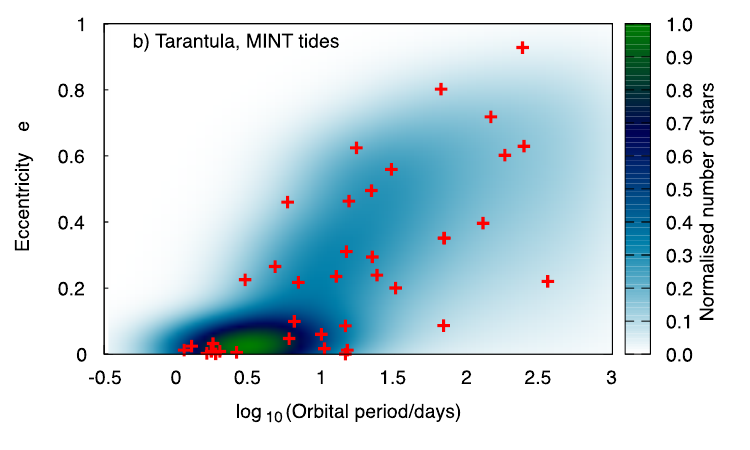}}    
    \caption{
    As Fig.~\ref{fig:m35} for the Tarantula cluster.
    }
    \label{fig:tarantula}
\end{figure*}

Lastly, we consider observations of a region populated by young, massive stars,
the Tarantula nebula.  This dense region of the Large Magellanic Cloud formed
through a series of star formation bursts 1 to 7 million years ago
\citep{schneider2018} and has a metallicity about half-solar corresponding to
$Z=0.008$ \citep{TD05,chowd15}.  We focus on the 38 O stars with orbital
properties from \citet{tar_almeida17}. We include stars in the mass range
$20-80 \,\Msun$ in our model population, we use mass ratios in the range
$0.5-1$ to match the observations, $Z=0.008$ and a reference age of 4\,Myr.  We
compare our model population with observations in the $e-\log_{10}(P/{\rm d})$
plane in Fig.~\ref{fig:tarantula}.

While it contains much younger and more massive stars than the previous
examples we present, this cluster follows the same statistical behaviour.  The
agreement is excellent as our model populations match the observations with a
distance below $0.2\sigma$.  The tidal prescription does not change this
agreement, which is to be expected as the cluster is very young and \bse and
\mint prescriptions start at a similar value, with \bse tidal coefficients
remaining constant but \mint dropping over time.

\subsection{Artificially modulating tides}
Despite having seen in section~\ref{sec:eqtide} that \mint tides are about ten
times as efficient as \bse in most solar-like stars, the statistics of the open
clusters seem to be dominated by the initial orbital distributions.  To measure
the effect of tides when relying on the Moe \& di Stefano distributions, we
modulate the efficiency of tides multiplying the orbital period and
eccentricity derivatives by a multiplicative strength factor. 

\begin{figure}
    \centerline{\includegraphics[width=.5\textwidth]{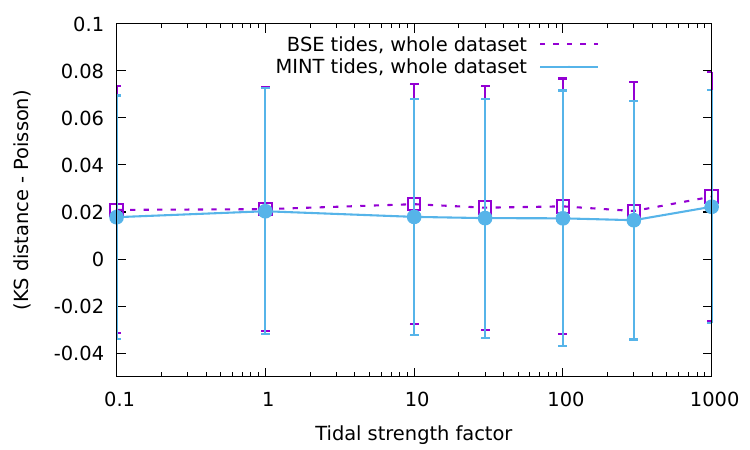}}
    \centerline{\includegraphics[width=.5\textwidth]{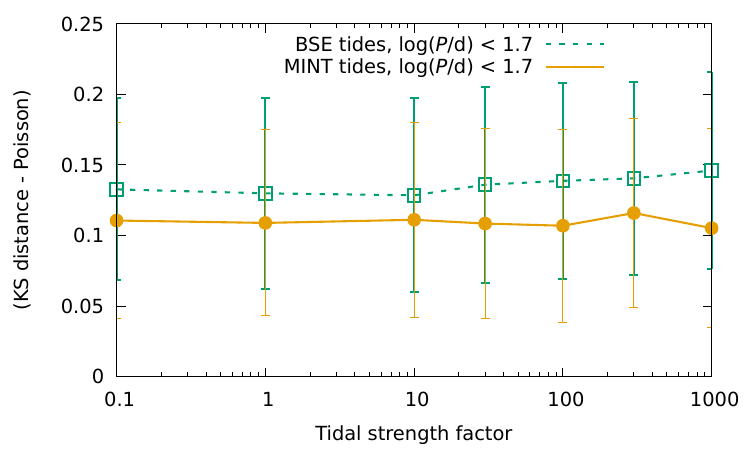}}
    \caption{
    Measure of the statistical agreement between M35 observations and
    populations computed with both \bse and \mint tides for various tidal
    strength factors, for the whole dataset (top) or a subset with $\log_{10}
    (P/\mathrm{d}) < 1.7$ (bottom). 
    }
    \label{fig:ts_M35trends}
\end{figure}

\begin{figure}
    \centerline{\includegraphics[width=.5\textwidth]{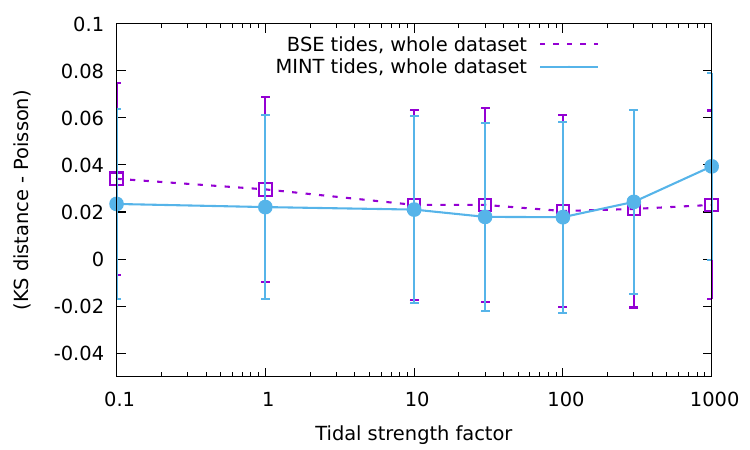}}
    \centerline{\includegraphics[width=.5\textwidth]{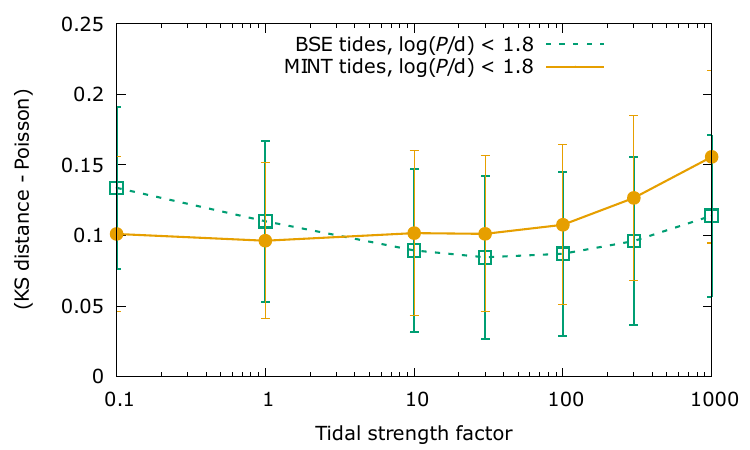}}
    \caption{
      As fig.~\ref{fig:ts_M35trends} for the M67 cluster, for the whole dataset (top) or a subset with $\log_{10}
    (P/\mathrm{d}) < 1.8$ (bottom).
    }
    \label{fig:ts_M67trends}
\end{figure}

To assess the impact of such a multiplicative change on the match with the
observations, we test strength factors from $0$ to $1000$. We perform this test
for the young cluster M35, and for the much older M67 that has the best-quality
data \citep{m67_G21}. We compare populations computed assuming Moe \& di
Stefano initial distributions with \bse and \mint tides, and compute the
statistics of both the whole dataset and the low-period subset of circularizing
systems.  Our results are summarized in Figs.~\ref{fig:ts_M35trends} and
\ref{fig:ts_M67trends} which show the statistical agreement between the model
populations and the observations.  

The comparison between the observations and the entire model population shows
that the two populations are compatible (at a distance of about $0.4\sigma$)
while the short-period systems lie at about $1.6-2.2\sigma$, matching the
numbers provided in Table~\ref{tab:stats_summary} for M35.  We see that this
agreement does not vary significantly despite the wide range of tidal strength
factors explored. This shows that main-sequence tides are not relevant to
justify current observations of binary systems in the M35 cluster and that the
choice of the initial distributions of period and eccentricity (that depend in
part on pre-main-sequence tidal dissipation) has a much greater impact.

M35 is a young open cluster (150Myr), while M67 is significantly older (4Gyr)
and is more likely to carry a tide signature.  As shown in
Fig.~\ref{fig:ts_M67trends}, the entire observed and model populations are
compatible (at a distance lower than $1\sigma$ for both tide prescriptions).
When focusing on circularizing systems at $\log_{10} (P/\mathrm{d}) < 1.8$, the
model populations lie $1.9$ and $1.6\sigma$ away from the observations when
using unaltered \bse and \mint tides, respectively. When using the detailed
implementation of Zahn's prescriptions with \mint, this agreement remains
roughly constant and worsens only when multiplying the base tidal dissipation
by more than 100. However, when the calculations are based on \bse simplified
prescriptions, we observe an improvement of the agreement between observations
and model by $\sim 0.4\sigma$ when multiplying tidal coefficients by 30 to 100.
While this improvement is noticeable only when focusing on low-period systems
and not significant, it matches the works of \citet{Belc08} and
\citet{Geller2013} that obtained more realistic circularization distributions
by multiplying \bse's convective damping by 50 to 100.

\section{Can synchronization help differentiate tidal prescriptions?}
\label{sec:poprot}
From the study of $e-\log_{10}(P/{\rm d})$ distributions of a variety of
clusters, it appears that circularization in stellar populations is dominated
by the initial distribution of eccentricities and periods, preventing us from
constraining tidal efficiency beyond the pre-main-sequence and
early-main-sequence phases.  However, tides do not only circularize binary
orbits, they also synchronize the stellar spins with the orbit over time. In a
priori eccentric systems, tides act more efficiently where the distance between
the stars is minimal, leading to a synchronization of spins with the orbit at
periastron. The resulting angular frequency is called pseudo-synchronous
\citep[][]{hut81}.  We study the evolution of stellar spins in open clusters in
search of a tide-dependent signature beyond the early main-sequence.

In this section we test \bse and \mint tidal prescriptions with Moe \& di
Stefano initial distributions focussing on the evolution of stellar rotational
properties.  As in section~\ref{sec:rotationrates}, we consider four initial
rotation settings: \bse's prescription from \citet{hurley00} given in
equation~(\ref{eq:vrot_bse}), a very low equatorial velocity of {$v_{\rm rot} =
10^{-4}$\,km s$^{-1}$}, initial breakup velocities or spin-orbit synchronous
rotation.  We focus here on the two clusters M35 and Tarantula, presented in
detail in section~\ref{sec:populations}.

\subsection{M35}
We start with our fiducial example, M35, assuming an initial rotation profile
matching the \bse prescription given in equation~(\ref{eq:vrot_bse}).

\begin{figure*}
    \centerline{
    \includegraphics[width=.5\textwidth]{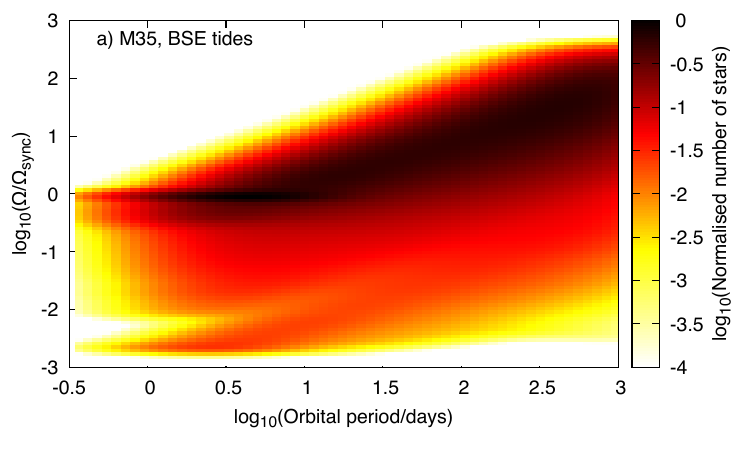}
    \includegraphics[width=.5\textwidth]{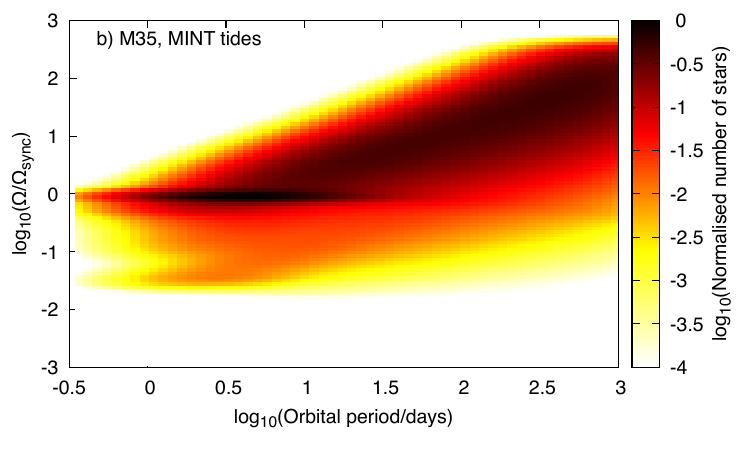}}    
    \caption{
    Angular frequency in units of the pseudo-synchronous angular frequency for
    M35 at age 150\,Myr, evolved with \bse (a) or \mint (b) tides starting from
    the \bse rotation velocities prescribed by equation~(\ref{eq:vrot_bse}).
    }
    \label{fig:M35omegasync}
\end{figure*}

\begin{figure*}
    \centerline{\includegraphics[width=.5\textwidth]{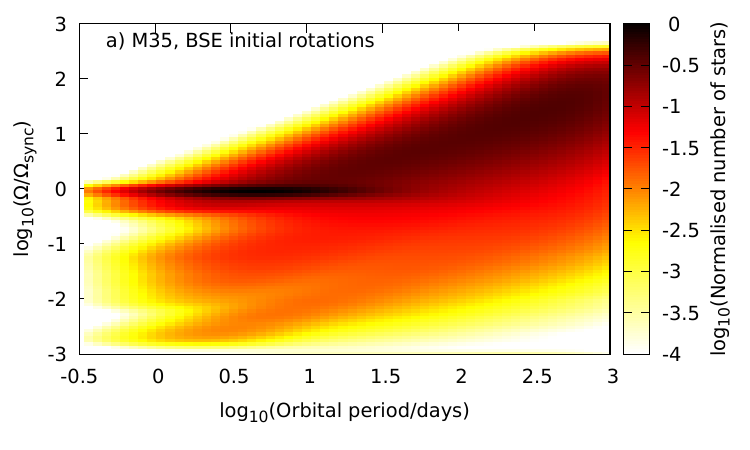}
    \includegraphics[width=.5\textwidth]{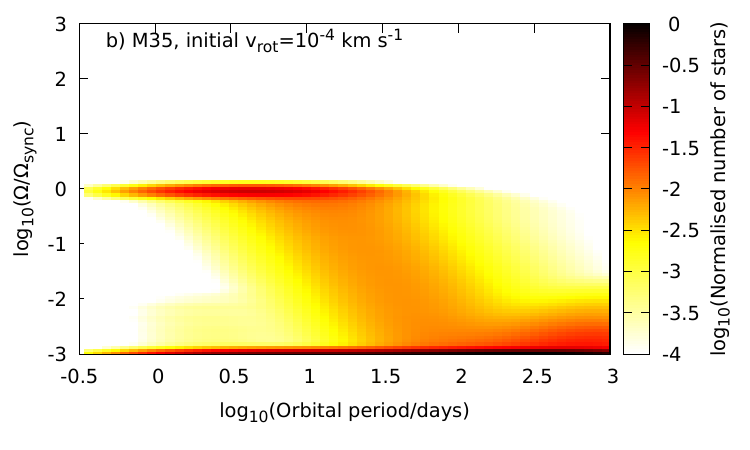}}
    \centerline{\includegraphics[width=.5\textwidth]{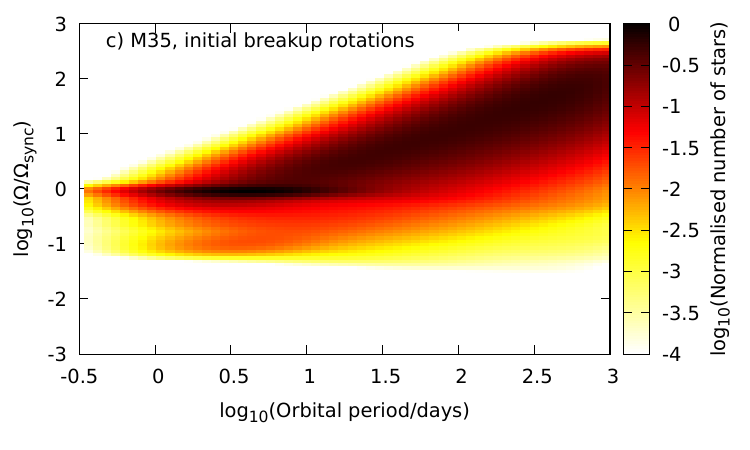}
    \includegraphics[width=.5\textwidth]{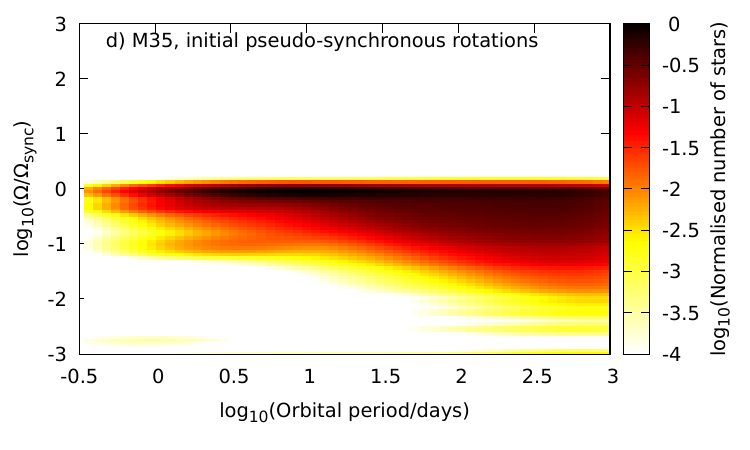}}    
    \caption{
    Angular frequency in units of the pseudo-synchronous angular frequency for
    M35 with \mint tides and Moe \& di Stefano initial distribution, assuming
    four different initial rotation profiles:
    (a) \bse rotation prescription,
    (b) $v_{\rm rot} =10^{-4}\,{\rm km}\ {\rm s}^{-1}$,
    (c) breakup velocity,
    (d) pseudo-synchronous rotation.
    }
    \label{fig:M35omegachange}
\end{figure*}

Fig.~\ref{fig:M35omegasync} presents the ratio of the stellar angular frequency
to the pseudo-synchronous one, on a logarithmic scale for both tidal
prescriptions.  In each panel, the high-count diagonal feature across the plot
is the signature of the initial rotation rate which is a function of mass only
while the pseudo-synchronous rate is a decreasing function of the orbital
period.  Stars in short-period systems are spun up by tides while those in
wider systems retain their initial low angular frequency.  This change in
behaviour happens at $\log_{10} (P/\mathrm{d}) \sim 1.5$ with both \bse and
\mint tides.  Tidal synchronization leads to a higher stellar count near
$\log_{10}(\Omega/\Omega_{\rm sync}) = 0$ for close-in systems. Such a feature
can be seen in both model populations, but is more prominent when using the
more efficient \mint tides.  This spin-up process activates in close-enough
systems owing to the highly non-linear dependence on $R/a$ in
equations~(\ref{eq:tsyn_conv}) and (\ref{eq:tsyn_rad}).  On the contrary, stars
in wide systems evolve towards slow rotation at all configurations of initial
rotation rates, even when they are initially set at breakup velocity on the
zero-age main sequence.  Angular momentum losses through magnetic braking slow
these stars in the first million years of their main-sequence evolution
regardless of the tidal prescription used.  This competition between magnetic
braking and tides is at the origin of the spread seen in
figures~\ref{fig:M35omegasync} and \ref{fig:M35omegachange}, and repeating this
experiment with other initial rotation prescriptions confirms this result, with
a dichotomy between spun up stars in short-period systems and slowly-rotating
wide systems.

Fig.~\ref{fig:M35omegachange} shows the model populations computed using \mint
tides for different initial rotation distributions.  Setting a
pseudo-synchronous angular frequency at the ZAMS, we would expect the ratio
$\Omega/\Omega_{\rm sync}$ to remain constant if only tides act on these stars,
but magnetic braking slows these stars and its competition with tides leads to
short-period systems near synchronicity in the range
$-0.2<\log_{10}\Omega/\Omega_{\rm sync}<0.2$, and wide systems rotating more
slowly and distributed over the wider range $-1.5<\log_{10}\Omega/\Omega_{\rm
sync}<0$.  Signatures of magnetic braking are also found in the sample starting
with $v_{\rm rot} =10^{-4}\,{\rm km}\ {\rm s}^{-1}$.  While some stars spin up
and reach synchronicity, about 80\% of them remain at very low rotation rates,
especially in wider orbits where tidal spin up is immediately compensated by
magnetic braking.  Similarly, systems forming at breakup velocity are rapidly
spun down by the combined effects of tides and magnetic braking so that most
signatures of the original high rotation rate vanish during the early cluster
evolution.

To quantify the effects of tides and their competition with magnetic braking,
we focus on close systems in Fig.~\ref{fig:M35omegachange} splitting them into
two bins. Retaining only close systems at ${\log_{10} (P/{\rm d)} < 1.5}$, we
count the fraction of stars in the $-0.2 < \log_{10} (\Omega/\Omega_{\rm sync})
< 0.2$ range that we deem synchronized. 
\begin{figure}
    \includegraphics[width=.5\textwidth]{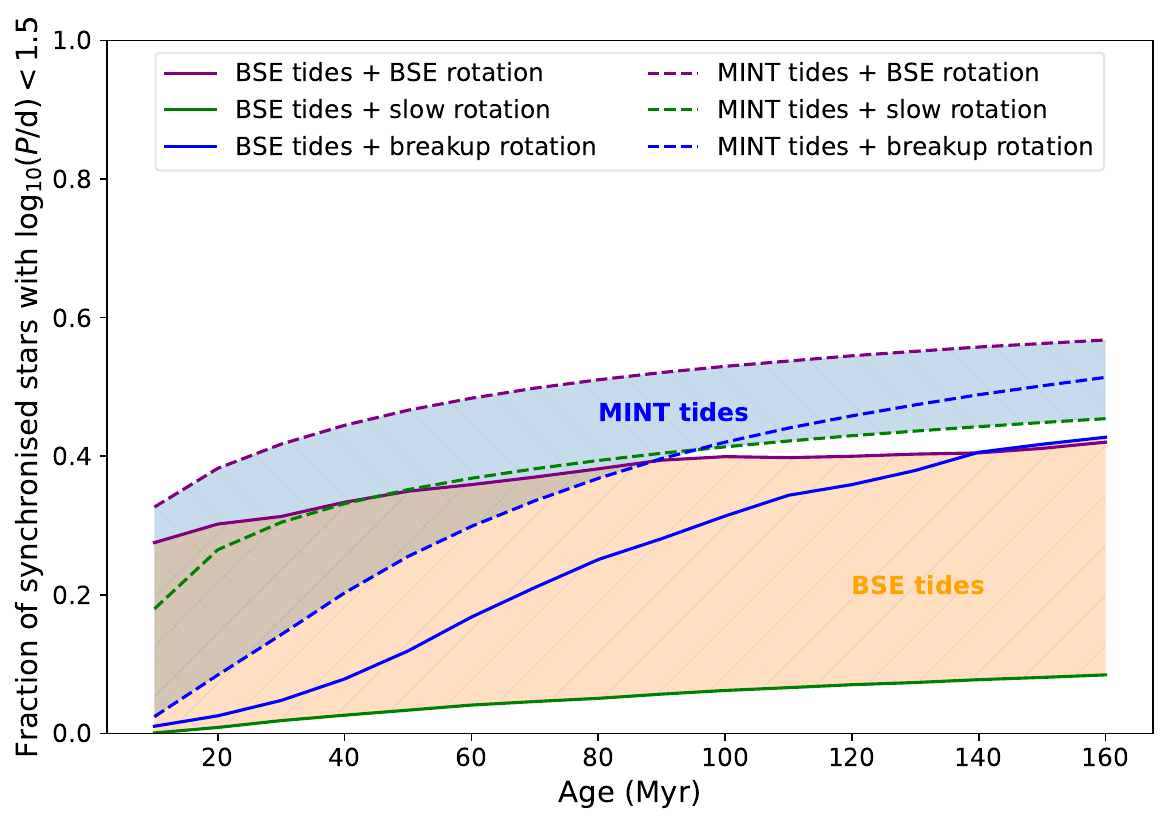}
    \caption{
    Fraction of M35 stars rotating at pseudo-synchronicity normalized to the
    total number of stars at $\log_{10} (P/\mathrm{d}) < 1.5$, assuming Moe \& di
    Stefano initial distributions.  Solid and dotted lines are obtained with
    \bse and \mint tides, respectively, for initial \bse rotation rates
    (purple), $v_{\rm rot} =10^{-4}\,{\rm km}\ {\rm s}^{-1}$ (green) and
    initial breakup rotations (blue).  The shaded areas show the domains of
    \mint (blue) and \bse (orange) tidal prescriptions.
    }
    \label{fig:M35ratiotosync}
\end{figure}

Fig.~\ref{fig:M35ratiotosync} shows the fraction of stars near
pseudo-synchronicity as a function of the population age, for $\log_{10}
(P/{\rm d}) < 1.5$. The age of M35 is estimated at 150\,Myr \citep[][]{MM05}.
These results show that \mint tides synchronize stellar spins with the orbit
faster and in more systems with respect to \bse tides, even when changing
between slow, breakup or \bse initial rotation rates. This is the result of the
higher efficiency of \mint equilibrium tides discussed in
section~\ref{sec:eqtide}. On average, we find that \mint equilibrium tides
predict 30 to 50\% pseudo-synchronized stars, while their \bse counterparts
predict on average one pseudo-synchronized star for each 5 that are not
synchronized. This difference provides a simple criterion that can be tested
through comprehensive surveys of clusters including solar-type binaries.
Including orbital parameters to determine the exact pseudo-synchronous rotation
period and individual stellar rotation periods would therefore allow us to
quantify the relative efficiency of tides and magnetic braking and favour a
prescription.  For instance, \citet{meibom_etal2006} use joint observations of
the orbital and rotational parameters of M35 systems and find that 2 of the 4
close systems they characterize are rotating synchronously. Such a result seems
to favour the \mint tidal prescription, but needs to be confirmed by more
systems in M35 and other clusters containing late-type binaries.

\subsection{Tarantula}
We repeat the above experiment with the Tarantula cluster, whose population of
young and massive O stars differs significantly from that of M35. Most
importantly, as these stars have a thick radiative envelope, they harbour
dynamical tides that follow the formalism laid out in section~\ref{sec:dyntide}
and angular momentum losses arise from stellar winds rather than magnetic
braking.  We use the wind mass loss prescription of \citet{schneider2018} that
was derived from observations of the Tarantula cluster.

\begin{figure*}
    \centerline{
    \includegraphics[width=.5\textwidth]{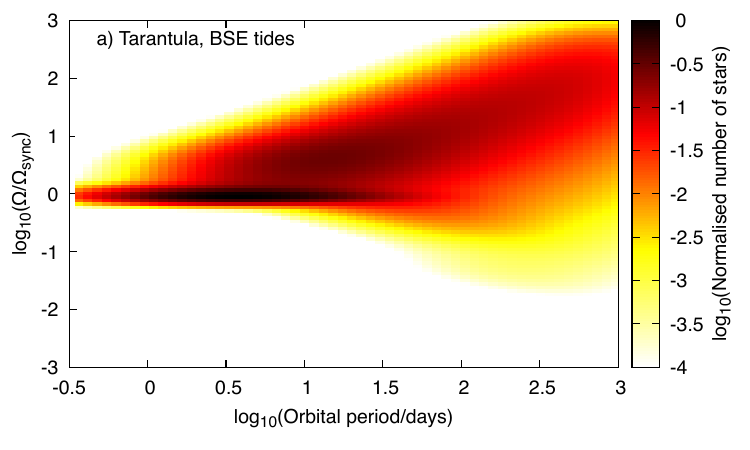}
    \includegraphics[width=.5\textwidth]{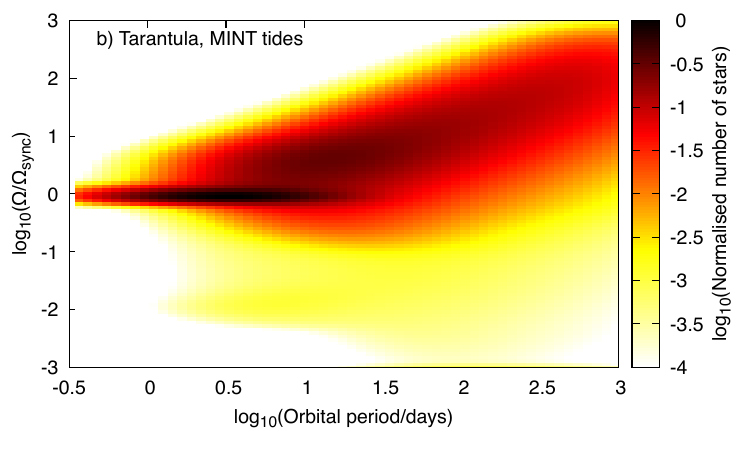}}    
    \caption{
    As Fig.~\ref{fig:M35omegasync} for the Tarantula population. 
    }
    \label{fig:Taromegasync}
\end{figure*}

In Fig.~\ref{fig:Taromegasync} we present the rotation rates in units of the
pseudo-synchronous rotation rate. As in Fig.~\ref{fig:M35omegasync}, the
diagonal feature at high periods is the signature of the initial rotation rate.
Systems with $\log_{10} (P/{\rm d}) < 1.5$ have a relatively high fraction of
pseudo-synchronized systems in the four cases shown here, that lies between
40\% for systems started at breakup and evolved with \mint tides and 60\% for
systems started at \bse rotation rates with \bse tides. 

\begin{figure}
    \includegraphics[width=.5\textwidth]{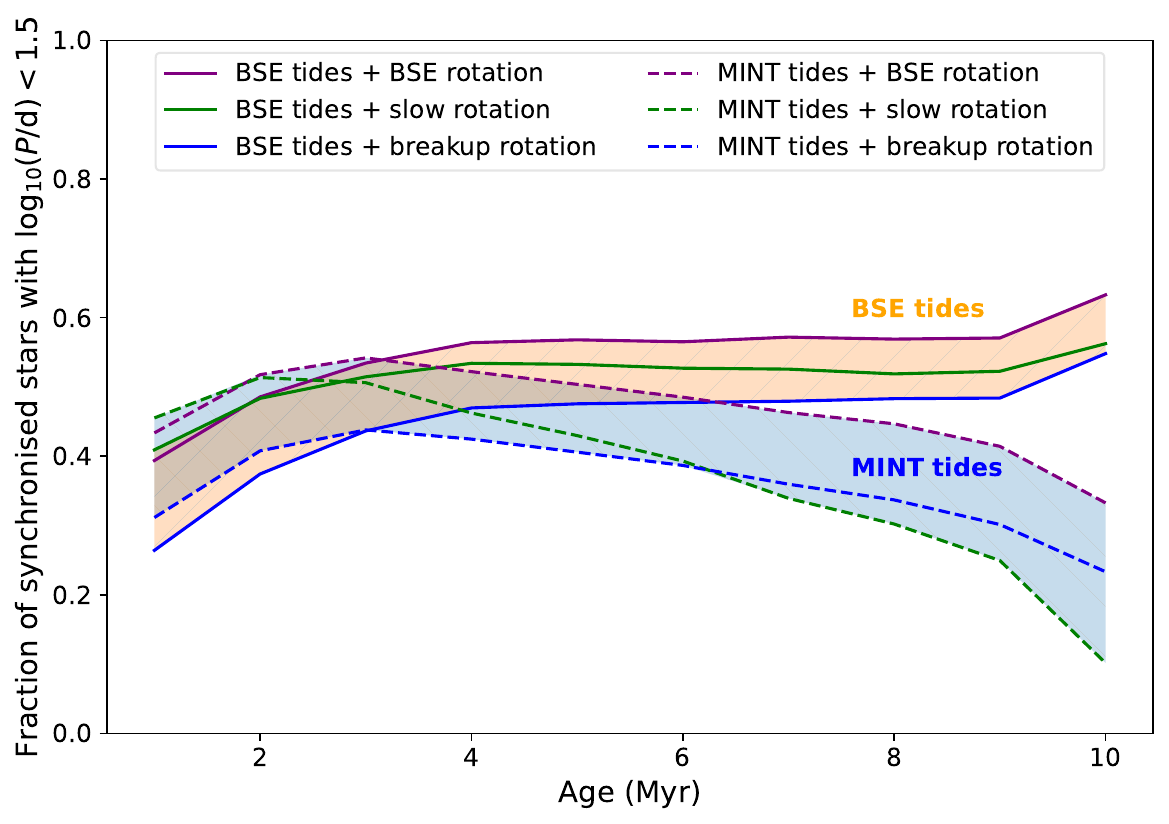}
    \caption{
    As Fig.~\ref{fig:M35ratiotosync} for Tarantula.
    }
    \label{fig:Tarratiotosync}
\end{figure}

Fig.~\ref{fig:Tarratiotosync} quantifies the evolution of this fraction of
synchronized stars as a function of age.  The Tarantula population formed
between 1 and 7\,Myr ago, with a peak of star formation 4\,Myr ago. At such
young ages, tides cannot be differentiated from synchronization processes, as
both tidal prescriptions have similar efficiencies near the ZAMS.  However,
\mint dynamical tides become less efficient over time while \bse dynamical
tides are not age-dependent.  Winds cause a loss of angular momentum for which
\mint dynamical tides cannot compensate after a certain age, so that some
systems fall out of pseudo-synchronicity, and the fraction of
pseudo-synchronous stars drops from $\sim 45$\% to $\sim 25$\%. On the
contrary, the model populations evolved with \bse tides have a steady $\sim
50$\% pseudo-synchronous stars.  As with equilibrium tides, this difference
induced by tidal prescriptions depends only slightly on the choice of initial
rotation, the range covered using different prescriptions is highlighted by the
shaded areas in Fig.~\ref{fig:Tarratiotosync}.  This would leave a detectable
signature in a 10 Myr old Tarantula twin cluster, as \bse dynamical tides would
predict as many pseudo-synchronized as non-synchronized systems while their
\mint counterparts predict only one pseudo-synchronized system for every three
that are not synchronized.  Appropriate measurements of the orbital and
rotational properties of close systems in older, massive-star open clusters can
thus decide which prescription is more suitable for dynamical tides between
\bse and \mint.

\section{Discussion}
\label{sec:discussion}
Our model populations show that circularization depends much more on the
initial orbital parameter distribution than on the tidal efficiency on the main
sequence (MS), even when using an ad hoc multiplicative factor, establishing
that MS tides are inefficient.  The presence of a short-period low-eccentricity
clump ($0.5<\log_{10}(P/{\rm d})<1.3, e<0.1$), surviving from the \citet{MS17}
initial orbital parameter distribution, confirms that pre-main-sequence (PMS)
interactions are crucial to describe the current eccentricity and period
distributions of observed open clusters.  Such a hypothesis was proposed by
\citet[]{zb89} and recent theoretical developments match our conclusions.
\citet[]{TM21} show, relying on the formalism of \citet{terquem21}, that
equilibrium tides are very efficient on the PMS but inefficient on most of the
MS.  It is only when stars develop an extensive convective envelope upon
reaching the very end of the MS or the subgiant phase (age $\gtrsim 10\,$Gyr
for a $1\,\Msun$ star) that their equilibrium tide efficiency increases to the
same order of magnitude as on the PMS.  Calculations invoking wave dissipation
through resonance-locking mechanisms usually yield increased tidal
circularization rates, which could lead to significant tides on the main
sequence. This is however not seen, as works such as \citet[]{ZW21} reach the
same conclusion that MS dynamical tides contribute much less than PMS tides to
circularization. An exhaustive implementation of these mechanisms over the
whole parameter range is necessary for population synthesis which would offer a
definitive answer.

The PMS tide efficiency is included in our calculations through the initial
distributions, that we take from \citet[]{MS17}. Further work by \citet[]{MK18}
investigates the origin of this distribution, and concludes that most of the
close binaries migrated to short periods during the PMS phase under the
associated action of the Kozai-Lidov mechanism (from a very long-period
triple), dynamical instability and tidal friction. Together, these formation
channels explain the large number of close binaries observed (highlighted by
the low-eccentricity short-period clumping in our model populations).  Our
calculations also show that circular and eccentric systems coexist at
intermediate periods ($3 - 20\,$days).  PMS migration explains this mixed
population with inflated stars on the Hayashi track circularizing efficiently
even at periods as long as a few weeks, and stars migrating later not
circularizing fully.  This situation would then remain generally the same
throughout the MS.  Investigating older populations, such as halo and field
stars with ages about $10\,$Gyr included in \citet[]{MM05}, would provide
insights on late-MS tidal dissipation. Recent developments in asteroseismology
and astrometry, ushered with the TESS and Gaia missions, offer unprecedented
statistics on binary systems in the field that can yield crucial insights on
tidal efficiency on and beyond the main sequence \citep{Beck2023}. However,
such populations are not as homogeneous as stellar clusters and their initial
conditions and ages would raise numerous uncertainties on the population
synthesis process.

Unfortunately, the relative inefficiency of MS tides renders the analysis of
circularization and the $e-\log_{10}(P/{\rm d})$ distribution a poor method of
constraining tides in clusters.  Defining a cut-off or circularization period
from the observed orbital parameters is a complicated task \citep[][]{MM05}
that might be irrelevant altogether.  \citet{Z22} may offer a solution to this
conundrum, by shifting the focus from circular to eccentric short-period
systems.  Based on the combined study of clusters presented here and Kepler
eclipsing binaries, they divide the samples into two populations:
nearly-circular binaries whose periods extend higher than measured
circularization periods and an envelope of eccentric systems at periods as low
as $\sim 3$ days. These populations also appear in the \citet{MS17} initial
distributions we use in this work.  Through a fit similar to the one performed
by \citet{MM05} to obtain circularization periods, but only applied to the most
eccentric systems at each orbital period, they derive the envelope period. This
indicator yields a statistically-significant difference between young (<1Gyr)
and old clusters (>3Gyr) and may carry the signature of MS equilibrium tides.
Another tentative explanation has been offered by \citet{bashi23}, that
analysed 17000 MS systems from the third Gaia data release, focussing on the
eccentric systems as well. They find that the envelope period scales linearly
with the stellar effective temperature rather than age, leading to a decreasing
envelope period with increasing stellar masses. While they highlight needed
observation advances, we contend population synthesis can offer theoretical
insights into the temperature dependence of tidal dissipation.  Studying the
impact of various tidal mechanisms on the cut-off periods estimated on circular
and eccentric systems by means of population synthesis codes will be the focus
of future work.  \\

In this work, we also propose the study of the rotational properties of cluster
stars, as synchronization carries the signature of tidal efficiency well into
the MS evolution of the stars in the system. We quantify this signature in
terms of the fraction of near-synchronous stars at short periods, which varies
with tidal efficiency and cluster age.  This criterion can be tested
observationally, by measuring both orbital parameters and individual stellar
spins through the combination of spectroscopy and photometry.  Early attempts
at such an analysis include \citet[][and references therein]{giuricin1984} who
find synchronization rates compatible with Zahn's theory. State-of-the-art
population studies that rely on modern stellar physics will be a key tool to
better constrain main-sequence tidal efficiency from surveys of rotational and
orbital parameters.  However such surveys are rare and sparse
\citep{meibom_etal2006, rebull2017}, and need to be completed and extended to
more clusters of main-sequence stars.  

The angular momentum changes of each star, and thus the fractions of
pseudo-synchronized rotators, are the result of the competition between
equilibrium tides and magnetic braking in low-mass stars or between dynamical
tides and stellar winds in massive stars.  Both winds and magnetic braking tend
to push stars out of synchronicity and explain why short-period systems can all
be circularized but still not synchronized with the orbit.  Investigating the
magnetic braking and wind mass-loss prescriptions in the literature and their
impact on the modelled fraction of stars rotating synchronously will also be
important to establish the measurability of tidal efficiency, and the topic of
future work.

The eccentricity-period distribution in open clusters is reminiscent of that of
barium/CH/CEMP-s stars. These stars are in binary systems and present the same
dichotomy between short-period circular systems and longer-period eccentric
systems that tidal interactions do not seem to explain \citep[][]{jorissen98,
jorissen16}.  The key to barium stars can be tides acting during the red-giant
phase.  The calculations we present here apply to other stages of stellar
evolution than the MS, and the inclusion of red giant stars in the \mint
evolution algorithm along with the relevant tides will be at the core of
upcoming work and is relevant to the study of numerous classes of stars.
Beyond barium stars, tides affect the fraction of synchronized systems and thus
the angular momentum budget available for Wolf-Rayet stars to form a soft-long
gamma-ray burst.  If dynamical tides cannot compensate for wind mass loss in
the late-MS phase and beyond, most massive stars will not evolve into a
collapsar that can form a disc necessary to the burst \citep[][]{izz04,
detmers08}.  Efficient dynamical tides are also necessary to form
chemically-homogeneous stars that provide a channel to binary black holes in
near-contact, low-metallicity massive binaries \citep{mandel16} while the
competition between tides and wind mass loss affects the number of mergers
predicted by this channel \citep{demink16}.  Both these applications require a
thorough study at low metallicity including post-MS evolution.

\section{Conclusions}
\label{sec:conclusions}
To summarize, we investigated the circularization process in open clusters, in
which two populations of binary systems coexist: circular systems with
$P<10-20$\,d and eccentric systems with $P>6-10$\,d, with both circular and
eccentric systems coexisting at intermediate periods in what appears to be a
tidally-driven transition period.  To investigate the origin of this
distribution, we implement and test detailed calculations of tidal dissipations
for main-sequence stars. We compute the coefficients $E$ and $E_2$ using Zahn's
theory of equilibrium and dynamical tides, relying on extensive grids of \mesa
structures (covering $M = 0.1 - 320\,\Msun$ and $Z=0 - 0.02$), and implement
them in the \binc stellar population code.  With respect to the ubiquitous \bse
prescriptions, the \mint implementation yields equilibrium tides 3 to 6 times
more efficient and dynamical tides similar at the ZAMS that then drop several
orders of magnitude with age. The impact on individual systems is significant.
The maximum period for circular systems at $1+0.5\,\Msun$ is 6 or 15 days with
\bse or \mint equilibrium tides respectively, for a $50+25\,\Msun$ system it is
25 days or 7.2 days with \bse or \mint tides respectively.

We then study $e-\log_{10}(P/{\rm d})$ distributions of binary stars in open
clusters over a wide range in age by modelling stellar populations with both
\bse and \mint tidal prescriptions and initial distributions derived from
bias-corrected observed properties \citep{MS17}.  We assess the agreement
between our model populations and orbital parameters measured for binary stars
in eight open clusters through a 2D Kolmogorov--Smirnov estimation. The
statistical agreement is excellent for most clusters, and mostly independent of
the tidal prescription used (both \mint and \bse tides typically lie within
$0.3\sigma$ of each other). This is due to a concentration of systems around
$\log_{10}(P/{\rm d})\sim 0.8,  e=0.05$, a direct consequence of the Moe \& di
Stefano distributions that tides do not modify over the main-sequence cluster
evolution.  This agreement does not change significantly even when multiplying
tides by a constant factor between 0 and 1000, but changing the initial
distributions to ones that do not include primordial short-period
low-eccentricity systems degrades the agreement very significantly for all
clusters.  We conclude that main-sequence tides have a very limited impact on
the statistical agreement between observations and model populations, which
makes the comparison between synthetic and observed $e-\log_{10}(P/{\rm d})$
diagrams an unsuitable way of constraining tidal prescriptions.

We then compute the synchronization of stellar spins with orbital periods and
find that \bse and \mint tides efficiencies consistently yield different
fractions of stars rotating at pseudo-synchronicity.  In clusters of low-mass
stars, \mint equilibrium tides are more efficient and lead to more synchronous
rotators over time while the situation is reversed in clusters of massive
stars. In M35 for instance, we expect about 40\% of the stars to rotate near
pseudo-synchronicity if \mint tides apply, while \bse tides would only yield
20\% of such stars.  For a massive-star cluster such as Tarantula, the fraction
of pseudo-synchronized O stars decreases with time as tides become less
efficient and wind mass loss removes angular momentum from the stars.  While
the synchronized rotator fraction is similar for both \bse and \mint tides in
Tarantula at its current age, a similar population at age 10\,Myr would have 3
times fewer synchronized stars if \mint tides apply in lieu of \bse tides.
These effects are significant and yield a workable criterion on the fraction of
stars rotating at pseudo-synchronicity that could be tested through combined
spectroscopic and photometric observations of the orbital parameters of the
systems and the individual stellar spins.

%%%%%%%%%%%%%%%%%%%%%%%%%%%%%%%%%%%%%%%%%%%%%%%%%%%%%%%%%%%%%%%%%%%%%%%%%%%%%%
\section*{Data availability}
The data underlying this article has been generated using free software and
will be shared upon request to the corresponding author.

\section*{Software}
We acknowledge the use of the following software:
\begin{itemize}
    \item The \mesa stellar evolution code
      (\href{http://mesa.sourceforge.net/}{http://mesa.sourceforge.net/}) and
      section \ref{sec:mesa}.
    \item The \binc stellar population synthesis code version 2.2.1, commit SHA
      \texttt{679b741fe},
      (\href{http://personal.ph.surrey.ac.uk/~ri0005/binary_c.html}{http://personal.ph.surrey.ac.uk/{\textasciitilde}ri0005/binary$\_$c.html})
      and section \ref{sec:binc}.
    \item The \textsc{binary\_c-python} software package \citep{joss}
    \item The Python implementation of the two-dimensional two-sample
      Kolmogorov--Smirnov estimator, by Zhaozhou Li
      (\href{https://github.com/syrte/ndtest}{https://github.com/syrte/ndtest}).
    \item The GNU Scientific Library \citep{gsl}.
\end{itemize}

\section*{Acknowledgements}
The authors acknowledge fruitful discussions during the PIMMS workshop
(\href{https://www.ias.surrey.ac.uk/event/pulsations-mass-stars/}{https://www.ias.surrey.ac.uk/event/pulsations-mass-stars/}).
We are grateful to the referee R. Mathieu for numerous suggestions that helped improve the paper greatly, to both him and
A. Nine for providing details about their observations, and to P. Das for her guidance about statistical inferences.  GMM and RGI acknowledge funding by
the STFC consolidated grants ST/L003910/1 and ST/R000603/1. DDH acknowledges funding by the UKRI grant H120341A.

\bibliographystyle{mnras}
\bibliography{bibliography} % if your bibtex file is called example.bib

\label{lastpage}
\clearpage
\appendix
%%%%%%%%%%%%%%%%%%%%%%%%%%%%%%%%%%%%%%%%%%%%%%%%%%%%%%%%%%%%%%%%%%%%%%%%%%%%%%

\section{Mathematical formalism}
\label{sec:math}

The impact of tides on orbital parameters is usually expressed in terms of
synchronization and circularization timescales.  The formalism in this section
is a summary of that of \citet{zahn77, hut81, zahn89} and \citet{SIDD13}.  All
equations presented in this section use masses, radii, and luminosities in
Solar units unless otherwise specified.  The dominating tides, and the
associated set of equations governing them, depend on whether the envelope of
the star is convective or radiative.

\subsection{Equilibrium tide: convective damping}
\subsubsection{Circularization and synchronization timescales}
\label{sec:eqtide}
In stars with a convective outer envelope, or fully-convective stars, we use the formalism of \citet{zahn89},
\begin{equation}
\label{eq:tcirc_conv}
    \frac{1}{\tcirc} = \frac{|\dot{e}|}{e} = 
    21\frac{\lambda_{10}}{\tau_{\rm conv}} \Tilde{q} \left(1+\Tilde{q}\right) \left( \frac{R}{a} \right)^{8}, 
\end{equation}
and
\begin{equation}
\label{eq:tsyn_conv}
    \frac{1}{\tsyn} = \frac{|\dot{\Omega}|}{\Omega - \omega} = 
    6\frac{\lambda_{22}}{\tau_{\rm conv}} \Tilde{q}^2  \frac{MR^2}{I} \left( \frac{R}{a} \right)^{6}.
\end{equation}

The overdot marks the time derivative, $a$ and $e$ are the semi-major axis and
eccentricity of the binary orbit, $\Tilde{q}$ is the ratio of the companion
mass to the mass of the star under consideration, $\omega$ is the orbit angular
velocity.  $M, R$ and $\Omega$ are the mass, radius and angular frequency of
the star under consideration, respectively, while $I$ is its momentum of
inertia. Finally, the convective turnover time $\tau_{\rm conv}$ is defined by
\citet{hurley02} as,
\begin{equation}
    \tau_{\rm conv} = 0.4311 \left[ \frac{M_{\rm env} R_{\rm env} \left( R - \frac{1}{2} R_{\rm env} \right)}{3L} \right]^{1/3}\ {\rm yr},
\label{eq:tau_conv}
\end{equation}
where $L$ is in turn the stellar luminosity, $M_{\rm env}$ the mass of the
convective envelope and $R_{\rm env}$ the depth of the core-envelope
boundary.\\ Note that this definition of $\tau_{\rm conv}$ is essentially
similar to that of~\citet{zahn77} but the convective turnover time is computed
assuming a typical convective element in the centre of the convective envelope
rather than at its base.

We present here the derivation of the $\lambda_{lm}$ coefficients, where the
indices $(l,m) = (0,1)$ and $(l,m) = (2,2)$ correspond to the spherical
harmonics used for the expansion of the tide-generating potential
\citep{zahn77, PS90} for circularization and synchronization respectively.
These coefficients are defined as
\begin{equation}
\label{eq:lambda2}
\lambda_{lm} = 0.8725 \left( \alpha' \right)^{4/3} E^{2/3} \times \mathcal{S}
\end{equation}
where $\alpha'$ is related to the mixing length $\alpha_{\rm MLT}$ through 
$\alpha' = 0.762 \alpha_{\rm MLT}$, the coefficient $E$ (not to be confused with $E_2$) depends on
the stellar structure at the core-envelope interface and $\mathcal{S}$ includes
dissipation via integrals on the stellar structure.  
We now detail the $E$ and  $\mathcal{S}$ terms.

\subsubsection{The $E$ coefficient}

\begin{figure}
    \includegraphics[width=.5\textwidth]{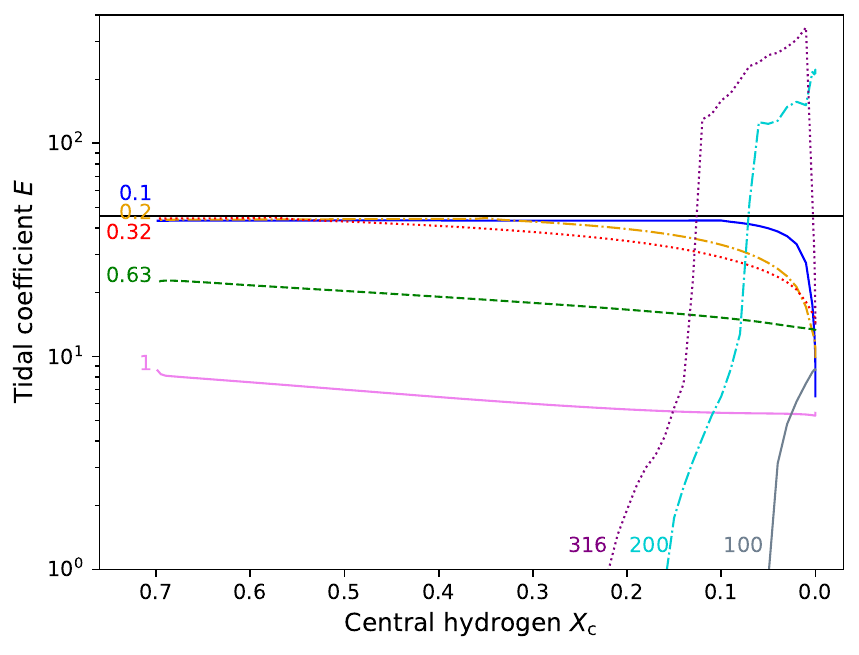} 
    \caption{
    The $E$ coefficient (equation~\ref{eq:E}) for a selection of stellar masses
    (indicated at the left or bottom of the plot), as a function of the central
    hydrogen mass fraction which is a proxy of age along the main sequence.
    The horizontal black line indicates the reference for a fully-convective
    star from \citet{zahn89}. 
    }
    \label{fig:Etracks}
\end{figure}

The structure parameter $E$ is defined as,
\begin{equation}
\label{eq:E}
    E = \rho_\mathrm{b} \frac{4\pi R^3}{M} \left( \frac{H_{P,\mathrm{b}} R}{r_\mathrm{b}^2} \frac{m_\mathrm{b}}{M} \right),
\end{equation}
where $H_P$ is the pressure scale height and the index ${\rm b}$ denotes the
base of the convective envelope. \citet{zahn89} provides the value $E=45.48$
as a maximum only reached in fully-convective stars.  Fig.~\ref{fig:Etracks}
shows $E$ computed from our \mesa models for a selection of masses (indicated
on the left or bottom of the plot). The solid black line is Zahn's value which
agrees with our early-main-sequence low-mass stars that are fully convective.
We also find that very massive stars, $M>90\,\Msun$, can have high $E$
coefficients making equilibrium tides efficient as massive stars develop an
extensive surface convection zone towards the end of the main sequence.  We
note that in these stars, $E$ is larger than the maximum value of
\citet{zahn89}.  In his analysis of polytropes, Zahn focused on low-mass
main-sequence and red giant stars, and did not consider such massive, inflated
main-sequence stars.

\citet{CC97} provide the prescription
\begin{equation}
\label{eq:E_claret}
    E = \frac{M_{\rm env}}{M} \times 
    \left[ \int_{x_\mathrm{b}}^{1} \left( \frac{2(1-x)}{5x} \right)^{3/2} x^2 {\rm d}x \right]^{-1},
\end{equation}
where $M_{\rm env}/M$ is the relative mass of the convective envelope and
$x=r/R$ is the relative radius throughout the star. Using this prescription will allow 
for comparison with our values of $\lambda_{lm}$ in paragraph \ref{sec:together}.

\subsubsection{The integrals underlying $\mathcal{S}$}
The $\mathcal{S}$ term is equation (\ref{eq:lambda2}) quantifies the tidal
efficiency through viscous dissipation in the convective envelope. For the first time in a
population code, we implement the derivation of \citet{zahn89} where
the viscous dissipation depends on the ratio between the tidal period
\begin{equation}
    \Pi_{lm} = \frac{2\pi}{|l\omega - m\Omega|},
\end{equation}
and the convective turnover time $\tau_{\rm conv}$ (equation~\ref{eq:tau_conv}).

If $\Pi_{lm} \geq 2 \tau_{\rm conv}$ 
throughout the convective envelope, $\lambda_{lm}$
depends on the stellar structure through,
\begin{equation}
\label{eq:S1}
    \mathcal{S} = \int_{x_{\rm b}}^{1} x^{22/3} (1-x)^2 {\rm d}x,
\end{equation}
where, $x = r/R$ is the reduced radius. Note that $\lambda_{lm}$ is, in this case, the
same for all indices $l$ and $m$, thus for both circularization and
synchronization. 

If the tidal and convective turnover timescales satisfy ${\Pi_{lm} = 2
\tau_{\rm conv}}$ at a depth $x_a$ in the convective envelope, the integral in
equation~(\ref{eq:S1}) splits into two terms, so that
\begin{equation}
\label{eq:S2}
\mathcal{S} = \int_{x_{\rm a}}^{1} x^{22/3} (1-x)^2 {\rm d}x 
                   +  x_{\rm a}^{7/6} (1-x_{\rm a})^{3/2} \int_{x_{\rm b}}^{x_{\rm a}} x^{37/6} (1-x)^{1/2} {\rm d}x.
\end{equation}
The first integral accounts for the viscous dissipation as in
equation~(\ref{eq:S1}) where ${\Pi_{lm} > 2 \tau_{\rm conv}}$. Where this
criterion is not valid, at $x_{\rm b} < x < x_{\rm a}$, convective cells cannot
travel their expected mean free path before a reversal of the tidal excitation,
which leads to a lower viscous dissipation that is accounted for by the second
integral in $\mathcal{S}$ \citep[][]{zahn89}.  The dependence on the tidal
period, and therefore the indices $l$ and $m$, is included in the integration
limit $x_a$. The existence and numerical value of $x_a$ is given by the roots
of the equation,
\begin{equation}
\label{eq:xa}
    x_{\rm a}^{7/6} (1-x_{\rm a})^{3/2} 
    = \left( \frac{5}{2} \right)^{3/2} \left( \alpha' \right)^{-2/3} E^{-1/3} \frac{\Pi_{lm}}{2 \tau_{\rm conv}}.
\end{equation}
The left-hand side of
equation~(\ref{eq:xa}) describes a bell-shaped function of $x_{\rm a}$ that
reaches its maximum, $c \sim 0.16$, at $x_{\rm a} = 7/16$.  The equation
has at most two roots, for the integral in equation~(\ref{eq:lambda2}) we only
retain the larger root if it is indeed inside the convective envelope (that is,
$x_{\rm a} > 7/16$ and $x_{\rm a} > x_{\rm b}$).  We thus find that such a root
exists if,
\begin{equation}
    \frac{\Pi_{lm}}{\tau_{\rm conv}}  \leq 2c \left( \frac{2}{5} \right)^{3/2} \left( \alpha'\right)^{2/3} E^{1/3}.
\end{equation}

We compute $\mathcal{S}$ from equation~(\ref{eq:S2}). Both integrals in
$\mathcal{S}$ have formal mathematical solutions that rely on the
hypergeometric function $_2 F_1$,
\begin{equation}
    \int_{x_{\rm min}}^{x_{\rm max}} x^a (1-x)^b {\rm d}x = 
    \left[ \frac{x^{a+1}}{a+1}  {}_2 F_1(a+1,-b ; a+2; x) \right]_{x_{\rm min}}^{x_{\rm max}}.
\end{equation}
In \binc, we compute the integrals numerically using the GNU Scientific Library
\citep[\textsc{GSL},][]{gsl} and they agree with the formal expression within
numerical precision.

\begin{figure}
   \includegraphics[width=.5\textwidth]{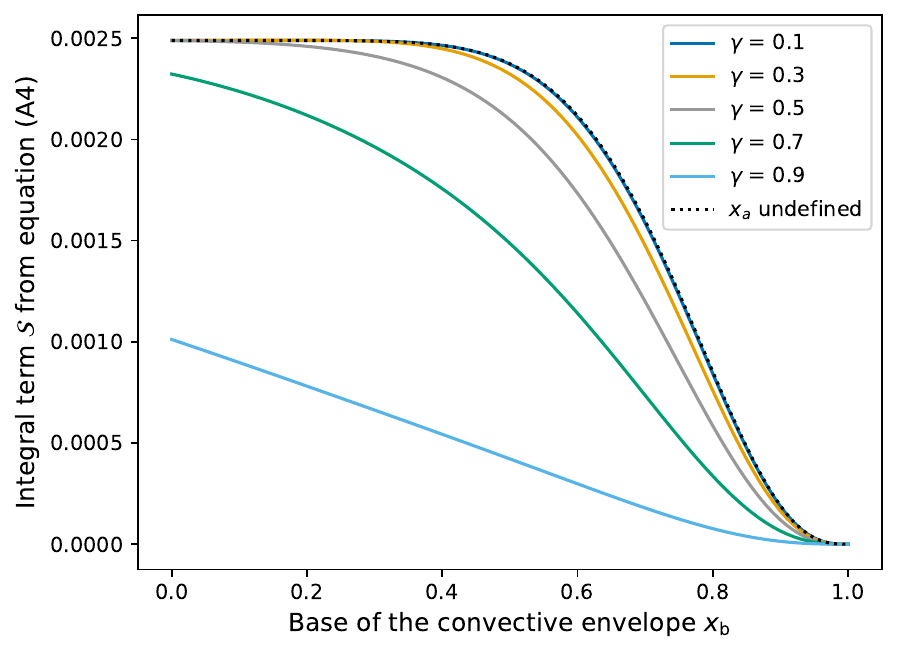} 
   \caption{
   $\mathcal{S}$ from equation~(\ref{eq:S2}) as a function of $x_{\rm b}$
   for various values of $\gamma$, defined by $x_{\rm a} = x_{\rm b} + \gamma
   (1-x_{\rm b})$. When $\gamma \rightarrow 0$, $x_{\rm a}$ becomes undefined
   and the corresponding $\mathcal{S}$ is marked by the dotted black line.
   }
   \label{fig:variousxa}
\end{figure}

Fig.~\ref{fig:variousxa} shows $\mathcal{S}$ at various convective interfaces
$x_{\rm b}$ and the critical depths $x_{\rm a}$.  In this figure, $x_{\rm a} =
x_{\rm b} + \gamma (1-x_{\rm b})$ where $\gamma$ is a constant between 0 and 1
that marks the limits and relative contributions of the integrals in
equation~(\ref{eq:S2}). $\gamma$ tends to 1 when the tidal period is shorter than
twice the convective turnover time everywhere in the star, leading to
$\lambda_{lm} = 0$, while $\gamma=0$ corresponds to tidal periods longer than
twice the turnover time anywhere in the star, in which case $x_{\rm a}$ is
formally undefined and equation~(\ref{eq:S2}) simplifies into
equation~(\ref{eq:S1}).  We find that $\mathcal{S} \rightarrow 0$ and thus that
{$\lambda_{lm} \rightarrow 0$} when the convective envelope is very thin
($x_{\rm b} \rightarrow 1$) or if the tidal period is comparable with the
convective turnover time throughout the envelope ($x_{\rm a} \rightarrow 1$).
The integral from $x_{\rm b}$ to $1$ dominates $\mathcal{S}$, as the integral
from $x_{\rm a}$ to $x_{\rm b}$ accounts for a reduced viscosity and thus
contributes less to the tidal torque.

\begin{figure}
   \includegraphics[width=.5\textwidth]{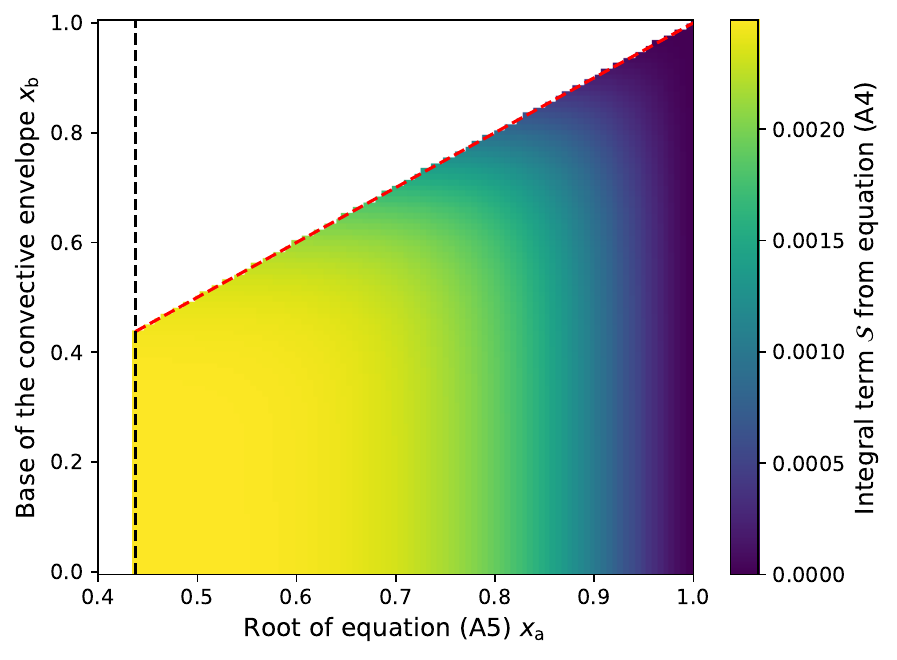} 
   \caption{
   $\mathcal{S}$ from equation~(\ref{eq:S2}) as a function of $x_{\rm b}$
   and $x_{\rm a}$.  $x_{\rm a}$ left of the black line $(x_{\rm a} = 7/16)$ or
   above the red line $(x_{\rm a} = x_{\rm b})$ are not physical.
   }   
   \label{fig:2dmap}
\end{figure}

\subsubsection{Putting equilibrium tides together}
\label{sec:together}
We have now detailed both $E$ and $\mathcal{S}$ and can use them in
equation~(\ref{eq:lambda2}).  The coefficients $\lambda_{lm}$ depend on the
ratio $\eta = 2 \tau_{\rm conv}/\Pi_{lm}$ through the limits of the integrals
underlying $\mathcal{S}$.  

\begin{figure}
    \includegraphics[width=.5\textwidth]{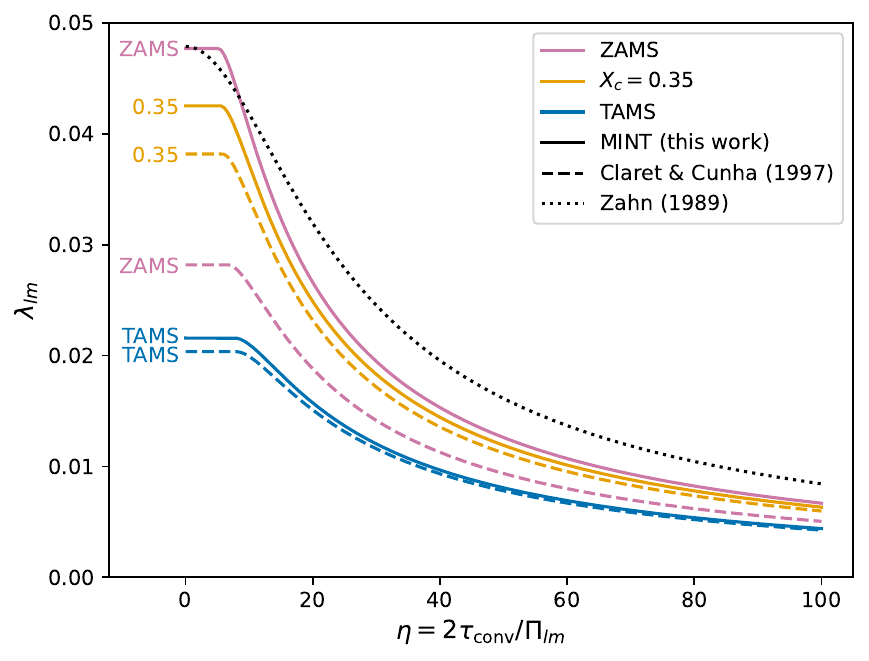} 
    \caption{
    Comparison between $\lambda_{lm}$ obtained from \mint (solid lines),
    \citet[dashed lines]{CC97} and \citet[dotted line]{zahn89} for a $0.32
    M_\odot$ star at the ZAMS (pink), halfway through the MS (orange) and at
    the TAMS (blue).  The \citet{zahn89} prescription is independent of age.
    }
    \label{fig:compare_lambda_0.32}
\end{figure}

\begin{figure}
    \includegraphics[width=.5\textwidth]{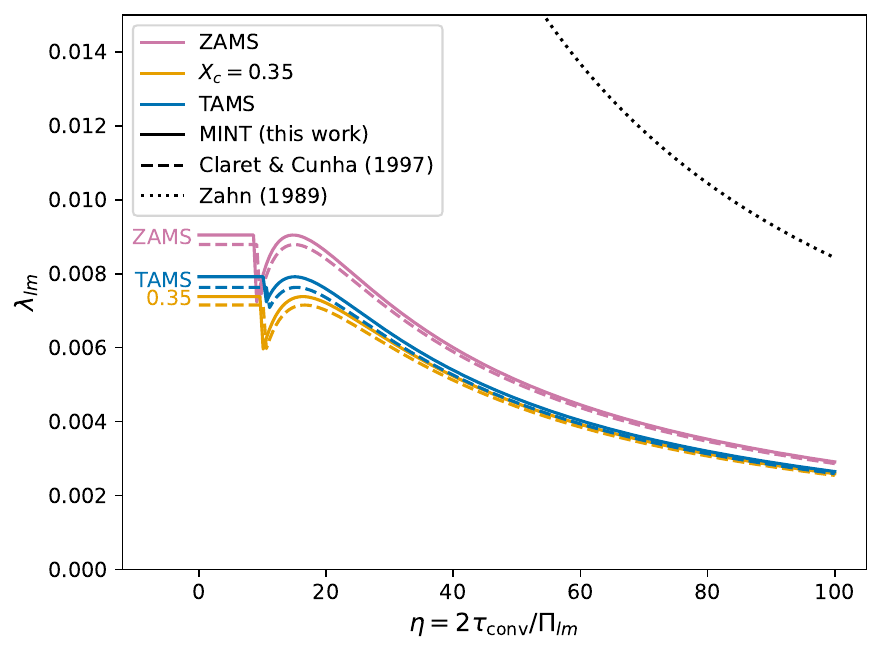} 
    \caption{As Fig.~\ref{fig:compare_lambda_0.32} for a $1 M_\odot$ star.}
    \label{fig:compare_lambda_1}
\end{figure}

In order to compare the \mint derivation with the literature, we use $E$ from
\citet{CC97} (equation~\ref{eq:E_claret}) and the fitting formula from
\citet{zahn89} :
\begin{equation}
  \lambda_{lm} = 0.019\ \alpha_{\rm MLT}^{4/3}\ \sqrt{\frac{320}{ 320+\eta^2} }.
\end{equation}
Figs.~\ref{fig:compare_lambda_0.32} and~\ref{fig:compare_lambda_1} present the
three calculations at $Z=0.02$, for $0.32$ and $1\,\Msun$ respectively. Each of
these plots presents $\lambda_{lm}$ at the beginning, the end, and halfway
through the main sequence ($X_c=0.35$).  As shown in
Fig.~\ref{fig:compare_lambda_0.32}, we find a good agreement between our
prescription and Zahn's fit for $M=0.32\,\Msun$ at the ZAMS as the star is
fully-convective there.  The agreement deteriorates at later ages as a
radiative core expands in the star.  On the contrary, the match between Claret
\& Cunha's prescription improves as the star ages and becomes more radiative.
Closer inspection of equation~(\ref{eq:E_claret}) provides an explanation. If
the star is fully convective, $M_{\rm env}/M \rightarrow 1$ and $x_{\rm
b}\rightarrow 0$, thus yielding $E = 20.13$. This is much smaller than both
Zahn's and \mint's $E$.  If the star is not fully convective, Zahn's
prescription becomes irrelevant while Claret \& Cunha's prescription relying on
the envelope mass yields a much better agreement with our full calculation.
This is visible in Fig.~\ref{fig:compare_lambda_1} for a $1\,\Msun$ star which
features a radiative core throughout its main-sequence evolution.\\


In \bse prescriptions, equations~(\ref{eq:tcirc_conv}) and (\ref{eq:tsyn_conv})
are cast in the slightly different form given by \citet{rasio96} to depend on
the parameter $(k/T)_{\rm c}$. This parameter encompasses the dependence of the
circularization and synchronization timescales on the ratio between convective
turnover time and tidal period in the same way as $\lambda_{22}$ and
$\lambda_{01}$.
Comparing the \bse and \mint mathematical derivations yields the equivalence,
\begin{equation}
    \lambda_{lm} = \frac{\tau_{\rm conv}}{2} \left( \frac{k}{T} \right)_{\rm c} 
                 = \frac{1}{21} \frac{M_{\rm env}}{M} \min\left(1, \eta^2 \right),
\label{eq:lambda_bse}
\end{equation}
where $\eta = 2\tau_{\rm conv}/\Pi_{lm}$ with $(l,m) = (2,2)$ 
for synchronization and $(l,m)=(0,1)$ for circularization, and 
$\tau_{\rm conv}$ is given by equation~(\ref{eq:tau_conv}). 

\begin{figure}
    \includegraphics[width=.5\textwidth]{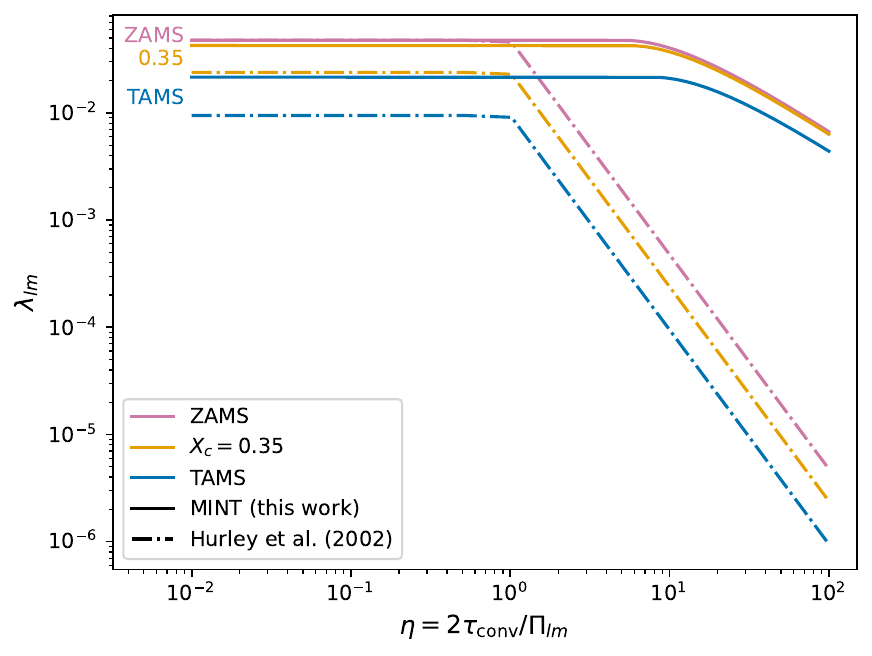} 
    \caption{Comparison between $\lambda_{lm}$ obtained from \mint 
    (solid lines) and \bse (dash-dotted lines) for a $0.32 M_\odot$ star
    at the ZAMS (pink), halfway through the MS (orange) and at the TAMS (blue).
    Both axes are in logarithmic scale.
    }
    \label{fig:lambda_log_log_0.32}
\end{figure}

\begin{figure}
    \includegraphics[width=.5\textwidth]{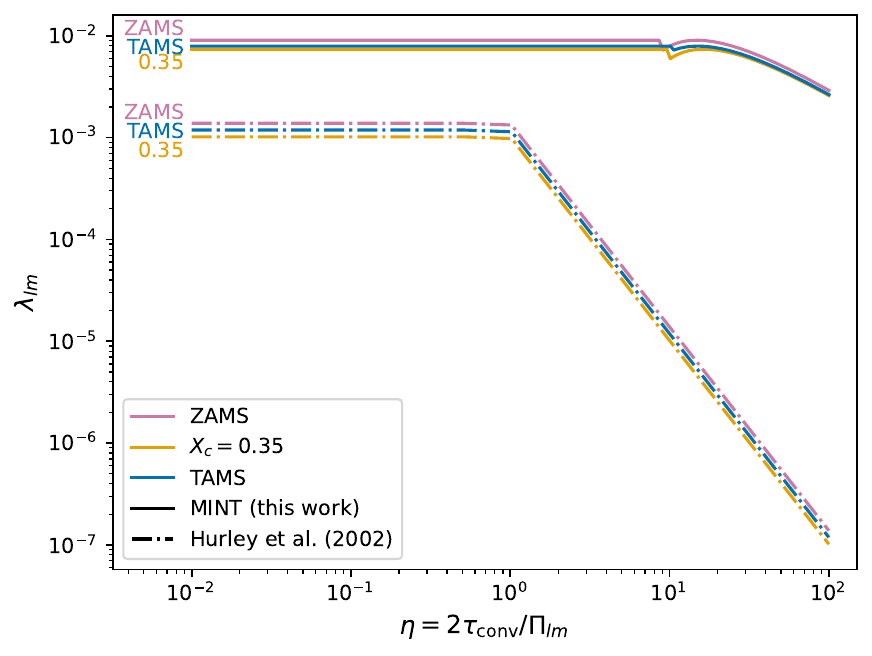} 
    \caption{As Fig.~\ref{fig:lambda_log_log_0.32} for a $1 M_\odot$ star.}
    \label{fig:lambda_log_log_1}
\end{figure}

We present in Figs.~\ref{fig:lambda_log_log_0.32}
and~\ref{fig:lambda_log_log_1} the comparison between our derivation and the
equivalent \bse $\lambda_{lm}$ obtained through equation~(\ref{eq:lambda_bse}).
These plots are in logarithmic scale to emphasize the asymptotic behaviour of
the coefficient at high and low $\eta$ ratios.  We see that \bse and \mint
calculations agree at low $\eta$ for the fully-convective $M=0.32\,\Msun$ model
(Fig.~\ref{fig:lambda_log_log_0.32}), and diverge when a radiative core builds
up. For the $M=1\,\Msun$ model (Fig.~\ref{fig:lambda_log_log_1}), the
disagreement reaches one order of magnitude and changes only slightly
throughout the main-sequence evolution, as the core properties are not changed
dramatically. This difference is intrinsic to the prescription used in
\citet{hurley02} that scales linearly with the radiative core mass, while our
estimate also takes the pressure scale height and core radius into account.  We
find that our \mint equilibrium tide prescriptions yield significantly higher
$\lambda_{lm}$ and thus faster circularization and synchronization.

The behaviour at large $\eta$ displayed in Figs.~\ref{fig:lambda_log_log_0.32}
and~\ref{fig:lambda_log_log_1} is striking.  The sharp drop in the \bse
calculation does not match the \mint asymptotic slower decrease, the \mint
$\lambda_{lm}$ is larger than \bse's by several orders of magnitude. Tidal
timescales derived from the \mint prescription are shorter than \bse's by the
same ratio.  This difference can be traced to the prescription used for the
viscosity when the convective turnover time and the tidal period are
comparable. As the tidal period becomes larger than twice the convective
turnover time ($\eta > 1$), convective cells cannot travel their expected mean
free path before a reversal of the tidal excitation.  This results in a drop in
dissipation at high $\eta$ that is parameterized either by multiplying the
viscosity by a factor $1/\eta$ \citep{zahn66} or by a factor $1/\eta^2$
\citep{GK77}.  Our derivation relies on the Zahn scaling which yields a
matching $-1$ slope at high $\eta$, while \bse relies on the Goldreich \&
Keeley model through the $ \min\left(1, \eta^2 \right)$ factor in
equation~(\ref{eq:lambda_bse}) leading to the steeper $-2$ slope in
Figs.~\ref{fig:lambda_log_log_0.32} and~\ref{fig:lambda_log_log_1}.  In
conclusion, our derivation of the equilibrium tide coefficients yields more
efficient tides over both the mass and tidal period range. This is confirmed in
section~\ref{sec:secular}.


%%%%%%%%%%%%%%%%%%%%%%%%%%%%%%%%%%%%%%%%%%%%%%%%%%%%%%%%%%%%%%%%%%%%%%%%%%%%%%%%%%%%%%%%%%%%%%%%%%%
\subsection{Dynamical tide: radiative damping}
\label{sec:dyntide}

\subsubsection{Circularization and synchronization timescales}
In stars with a radiative envelope, the circularization and synchronization timescales 
$\tcirc$ and $\tsyn$ are given by \citet{zahn75,zahn77}:
\begin{equation}
\label{eq:tcirc_rad}
    \frac{1}{\tcirc} = \frac{|\dot{e}|}{e} = 
    \frac{21}{2} \left( \frac{GM}{R^3} \right)^{1/2} \Tilde{q} \left(1+\Tilde{q}\right)^{11/6}
    E_2 \left( \frac{R}{a} \right)^{21/2}, 
\end{equation}
and
\begin{equation}
\label{eq:tsyn_rad}
    \frac{1}{\tsyn} = \frac{|\dot{\Omega}|}{\Omega - \omega} = 
    5 \left( 2^{5/3} \right) 
    \left( \frac{GM}{R^3} \right)^{1/2} \Tilde{q}^2 \left(1+\Tilde{q}\right)^{5/6}
    \frac{MR^2}{I} E_2 \left( \frac{R}{a} \right)^{17/2},
\end{equation}
where most parameters follow the definitions given in
equations~(\ref{eq:tcirc_conv}) and~(\ref{eq:tsyn_conv}).  The key parameter is
the remaining quantity $E_2$ (not to be confused with $E$) that depends on
integrals of the stellar structure. 

\subsubsection{The coefficient $E_2$}
In the \bse implementation, $E_2$ is related to the stellar mass $M$ through a
fit to data provided by \citet{zahn75},
\begin{equation}
E_2 = 1.592\times 10^{-9} M^{2.84}.
\end{equation}
Most notably, this relation is age- and metallicity-independent.  There have
been several attempts at including the age-dependent core recession in the
calculation of $E_2$, developed by \citet{zahn77}, \citet{CC97} and
\citet{SIDD13}. This latter work provides an attempt at computing $E_2$ as a
function of mass, age, and metallicity.  Our detailed derivation follows theirs
which we summarize here.  

\begin{equation}
\label{eq:E2}
    E_2 = \gamma_2 \frac{\rho_{\rm b} R^3}{M} 
          \left[ \frac{R}{g_{\rm s}} \frac{\rm d}{{\rm d}x}\left( \frac{-gB}{x^2} \right)_{\rm b} \right]^{-1/3}
           \left( H_2 \right)^2
\end{equation}
where the index ${\rm b}$ labels the convective core boundary and ${\rm s}$ the
stellar surface, $x$ is the relative radius and $g$ the local gravity.  The
constant $\gamma_2$ is,
\begin{equation}
    \gamma_2 = \frac{3^{8/3} \left[ \Gamma(4/3)\right]^2 }{5\cdot 6^4/3} \sim 0.27384... ,
\end{equation}
and $B$ is defined from stellar structure quantities through
\begin{equation}
    B = \frac{\rm d}{{\rm d}r} \ln\rho - \frac{1}{\Gamma_1} \frac{\rm d}{{\rm d}r} \ln P 
      = \frac{\delta}{H_P} \left( \nabla - \nabla_{\rm ad} - \frac{\psi}{\delta}\nabla_\mu\right).
\end{equation}
$B$ is related to the Brunt-V\"ais\"al\"a frequency $N$ through ${N^2 = -gB}$.
The Brunt-V\"ais\"al\"a frequency is known to vary sharply at the interface
between convective and radiative zones in upper-main-sequence stars. Assessing
the exact location of the core boundary is thus key for the calculation of
$E_2$.

The last term remaining in the calculation of $E_2$ is $H_2$, defined as,
\begin{equation}
    H_2 = \frac{1}{X(x_{\rm b}) Y(1)} \int_{0}^{x_{\rm b}} X \left[ \frac{{\rm d}^2 Y}{{\rm d}x^2} - \frac{6Y}{x^2}\right] {\rm d}x, 
\end{equation}
where $X$ and $Y$ are the solutions of the differential equations
\begin{equation}
    \frac{{\rm d}^2 X}{{\rm d}x^2} 
    - \frac{{\rm d}\ln\rho}{{\rm d}x} \frac{{\rm d}X}{{\rm d}x} - 
    \frac{6}{x^2}X = 0,
\end{equation}
and
\begin{equation}
    \frac{{\rm d}^2 Y}{{\rm d}x^2} 
        - \frac{6}{x}\left( 1 - \frac{\rho}{\Bar{\rho}} \right) \frac{{\rm d}Y}{{\rm d}x} 
        - \frac{6}{x^2} \left( 2\frac{\rho}{\Bar{\rho}} -1 \right)Y = 0,
\end{equation}
in which $\Bar{\rho} = 3 m/ (4 \pi r^3)$ is the density of the material included
inside radius $r$.  The solutions of these equations are computed
numerically following the method presented in Appendix B of \citet{SIDD13}.

\subsubsection{Comparison with prescriptions in the literature}
\begin{figure}
    \includegraphics[width=.5\textwidth]{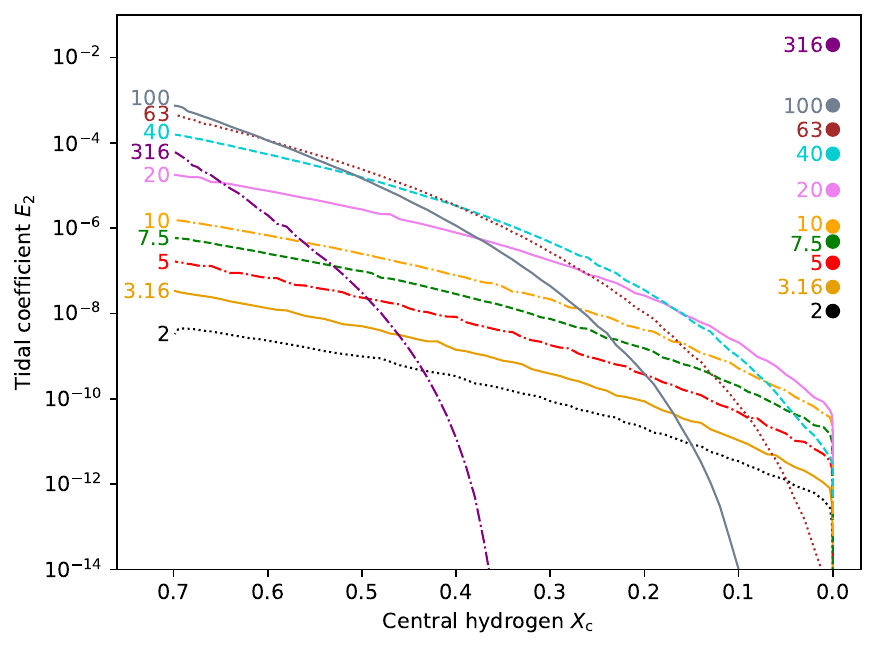} 
    \caption{
    The $E_2$ coefficient (equation~\ref{eq:E2}) for a selection of stellar
    masses (indicated on the left of the plot), as a function of the central
    hydrogen mass fraction at $Z=0.02$.  The dots in the upper-right corner
    give the age- and metallicity-independent value from \bse, based on a fit
    of \citet{zahn75}.
    }
    \label{fig:E2tracks}
\end{figure}

Within the \mint framework, we estimate each of the terms contributing to $E_2$
by evaluating the necessary structure quantities and integrals throughout grids
of \mesa models. We derive accurate $E_2$ for a wide range of masses, ages, and
metallicities. These are shown in Fig.~\ref{fig:E2tracks} for masses from $2$
to $316\,\Msun$ at $Z=0.02$. These models all feature a convective core
surrounded by an extensive radiative zone in which low-frequency gravity waves
dissipate energy.  Points in the upper-right corner of the plot show the
age-independent \bse $E_2$ values \citep{hurley02} for the same selection of
masses.  We note that the \bse prescription generally over-estimates $E_2$. It
roughly matches \mint at the zero-age main sequence, but as evolution proceeds
and the stellar convective core recedes, the \mint $E_2$ decreases by several
orders of magnitude.  The most massive stars see their structure change on the
main sequence, starting with a convective core and a radiative surface, then
developing a convective envelope as they inflate towards the end of the main
sequence.  This leads to a drop in $E_2$ in the late main sequence matching the
timing of the increase in $E$ highlighted in Fig.~\ref{fig:Etracks}. In those
stars, the main tidal dissipation mechanism shifts from radiative to convective
damping.

This is also an improvement on \citet{SIDD13}, as their models yielded
numerically noisy values and the corresponding prescription was expressed in
the form of a set of fitting formulae of the mass and the metallicity.

Other prescriptions for $E_2$ found in the literature rely on the radius of the
convective core, through scaling relations of the form,
\begin{equation}
    E_2 = 10^{a} \left( \frac{R_{\rm conv}}{R} \right)^b,
\end{equation}
where \citet{yoon2010} recommend ${a=-1.37}, {b=8}$, while \citet{qin2018} use
${a=-0.42}, {b=7.5}$ for hydrogen-rich stars.

\begin{figure}
    \includegraphics[width=.5\textwidth]{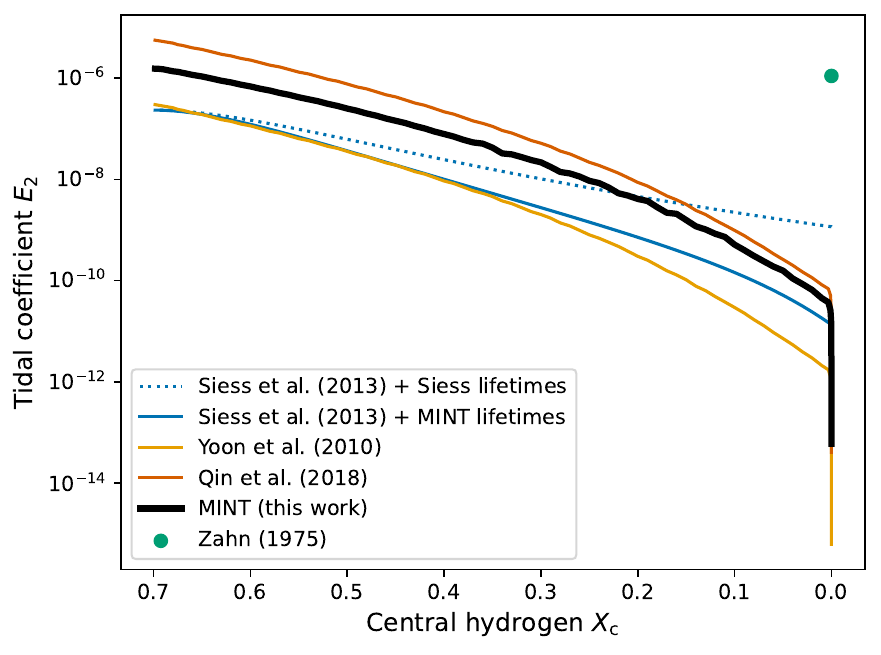} 
    \caption{Comparison between coefficients $E_2$ 
    for a $10 M_\odot$ star. 
    Colours mark the various prescriptions. Two calculations using Siess's
    prescriptions are presented (see main text) while the Zahn prescription
    yields a constant value shown as a dot in the top-right corner.
    }
    \label{fig:compare_E2_10}
\end{figure}

Fig.~\ref{fig:compare_E2_10} presents a comparison between these prescriptions
and our derivation for a $10\,\Msun$ star.  While the prescription from
\citet{zahn89} matches our derivation of $E_2$ only at the ZAMS, other
prescriptions all offer an agreement within one order of magnitude. Most
notably, prescriptions relying on the convective core radius follow the same
trend as our complete calculation and differ roughly by a multiplicative
constant.  Fig.~\ref{fig:compare_E2_10} also shows two calculations relying on
\citet{SIDD13}: the dotted blue line relies on their fitting relation for the
main-sequence lifetime (upon which $E_2$ strongly depends), while the solid
blue line relies on the lifetimes obtained from our \mesa runs. The difference
in the lifetimes emerges from the use of two different mixing-length parameters
in the models ($\alpha_{\rm MLT} =1.75$ in the models of \citealt{SIDD13} and
$2$ in ours), along with updated equations of state and opacity tables in the
\mesa code on which our calculations rely. 

\section{Population studies for a range of open clusters}
\label{sec:other_clusters}
In this appendix, we present the $e-\log_{10}(P/{\rm d})$ diagrams and
statistics associated with the open clusters studied in
section~\ref{sec:populations}. 

%%%%%%%%%%%%%%%%%%%%%%%%%%%%%%%%%%%%%%%%%%%%%%%%%%%%%%%%%%%%%%%%%%%%%%%%%%%%%%%%%%%%%%%%%%%
\begin{figure*}
    \centerline{
    \includegraphics[width=.5\textwidth]{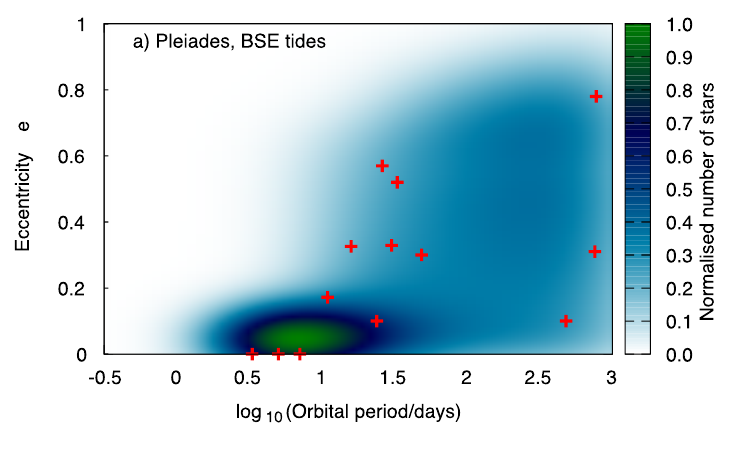}
    \includegraphics[width=.5\textwidth]{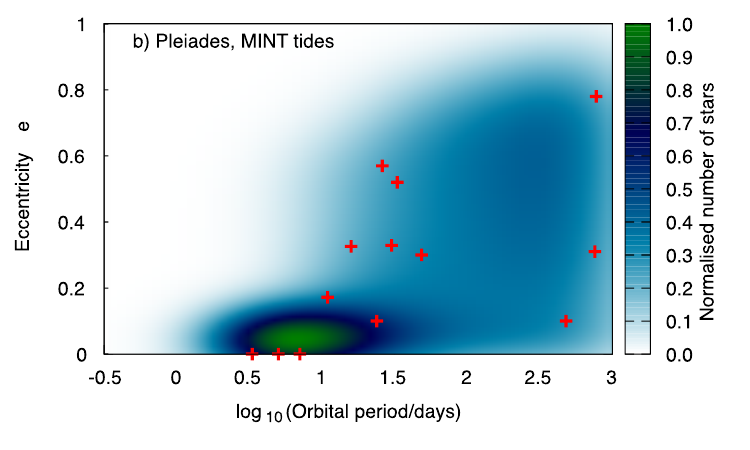}}    
    \caption{
    Comparison between Pleiades observations (red crosses) and the stellar
    counts calculated populations at the documented cluster age, normalized at
    the highest bin count (colour map). Starting from \citet{MS17} initial
    distributions, we test both \bse (a) and \mint (b) tidal prescriptions. 
    }
    \label{fig:pleiades}
\end{figure*}

%%%%%%%%%%%%%%%%%%%%%%%%%%%%%%%%%%%%%%%%%%%%%%%%%%%%%%%%%%%%%%%%%%%%%%%%%%%%%%%%%%%%%%%%%%%
\begin{figure*}
    \centerline{
    \includegraphics[width=.5\textwidth]{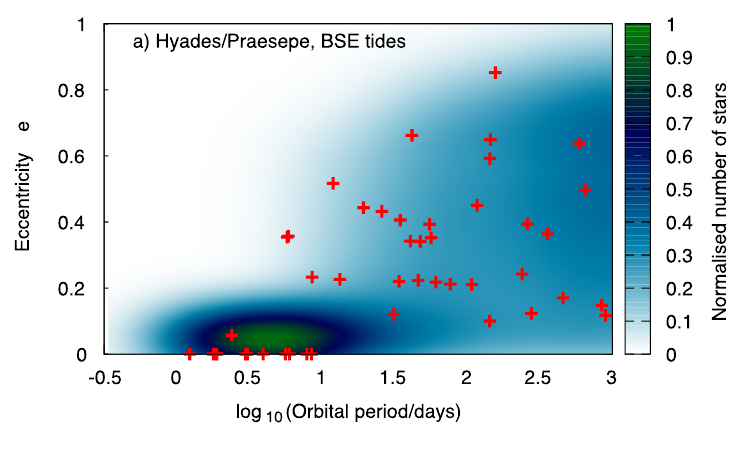}
    \includegraphics[width=.5\textwidth]{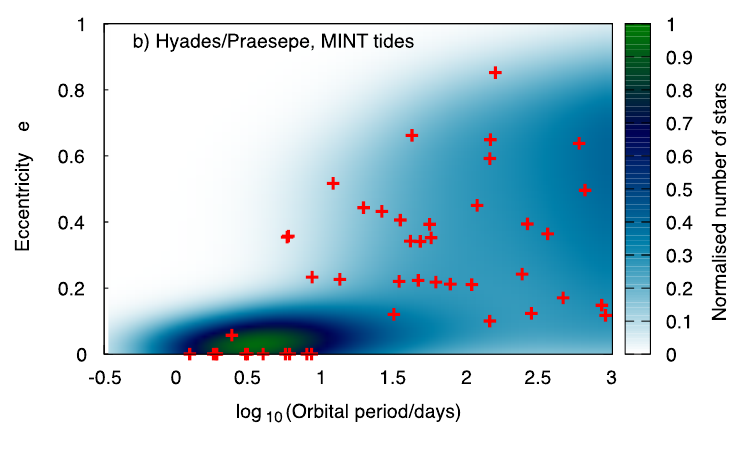}}    
    \caption{
    As Fig.~\ref{fig:pleiades} for the Hyades and Praesepe twin clusters.
    }
    \label{fig:hypr}
\end{figure*}

%%%%%%%%%%%%%%%%%%%%%%%%%%%%%%%%%%%%%%%%%%%%%%%%%%%%%%%%%%%%%%%%%%%%%%%%%%%%%%%%%%%%%%%%%%%
\begin{figure*}
    \centerline{
    \includegraphics[width=.5\textwidth]{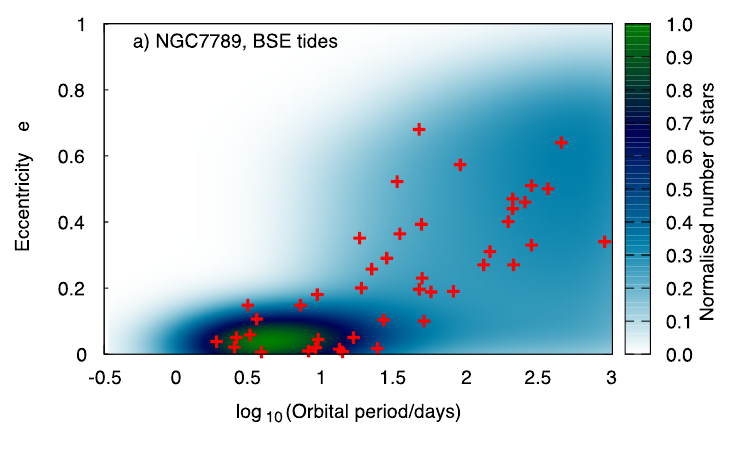}
    \includegraphics[width=.5\textwidth]{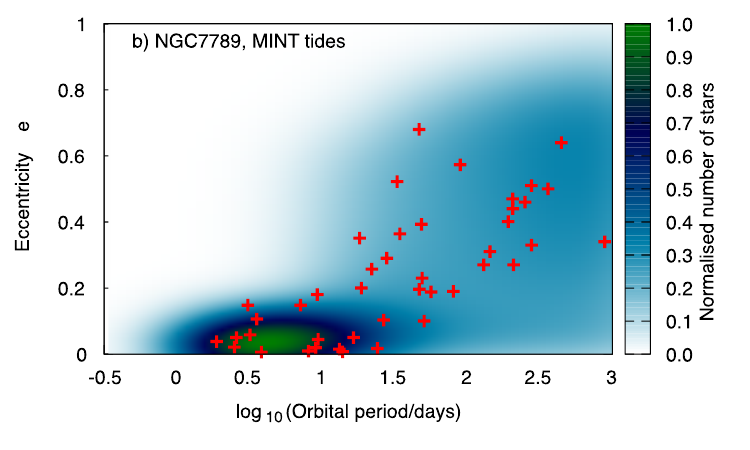}}
    \caption{
    As Fig.~\ref{fig:pleiades} for the NGC 7789 cluster.
    }
    \label{fig:ngc7789}
\end{figure*}

%%%%%%%%%%%%%%%%%%%%%%%%%%%%%%%%%%%%%%%%%%%%%%%%%%%%%%%%%%%%%%%%%%%%%%%%%%%%%%%%%%%%%%%%%%%
\begin{figure*}
    \centerline{
    \includegraphics[width=.5\textwidth]{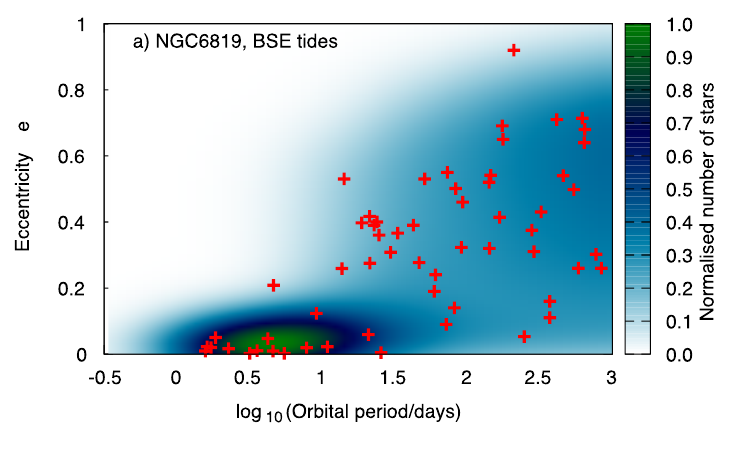}
    \includegraphics[width=.5\textwidth]{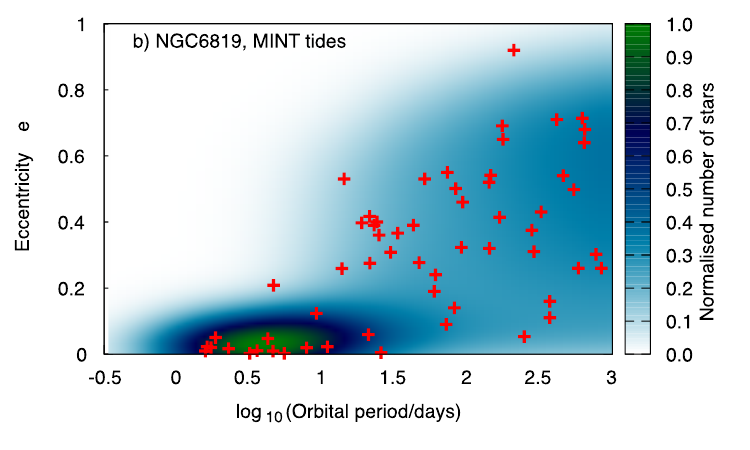}}
    \caption{
    As Fig.~\ref{fig:pleiades} for the NGC 6819 cluster.
    }
    \label{fig:ngc6819}
\end{figure*}

%%%%%%%%%%%%%%%%%%%%%%%%%%%%%%%%%%%%%%%%%%%%%%%%%%%%%%%%%%%%%%%%%%%%%%%%%%%%%%%%%%%%%%%%%%%
\begin{figure*}
    \centerline{
    \includegraphics[width=.5\textwidth]{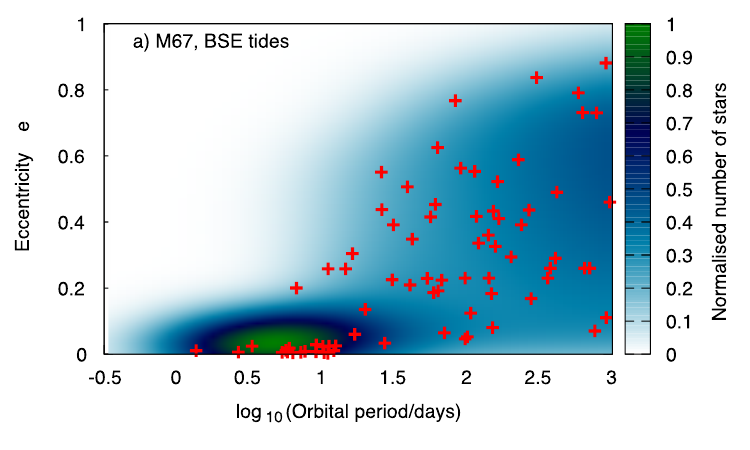}
    \includegraphics[width=.5\textwidth]{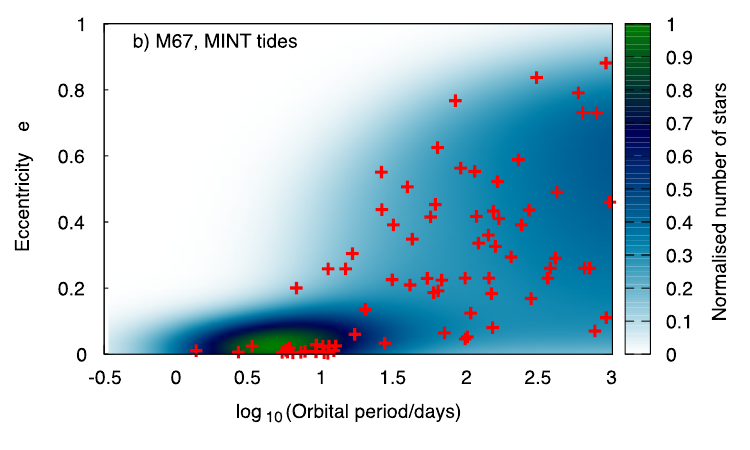}}
    \caption{
    As Fig.~\ref{fig:pleiades} for the M67 cluster.
    }
    \label{fig:M67}
\end{figure*}

%%%%%%%%%%%%%%%%%%%%%%%%%%%%%%%%%%%%%%%%%%%%%%%%%%%%%%%%%%%%%%%%%%%%%%%%%%%%%%%%%%%%%%%%%%%
\begin{figure*}
    \centerline{
    \includegraphics[width=.5\textwidth]{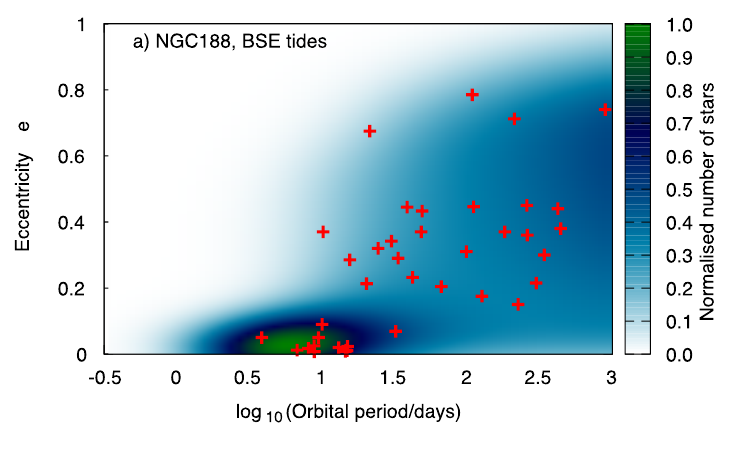}
    \includegraphics[width=.5\textwidth]{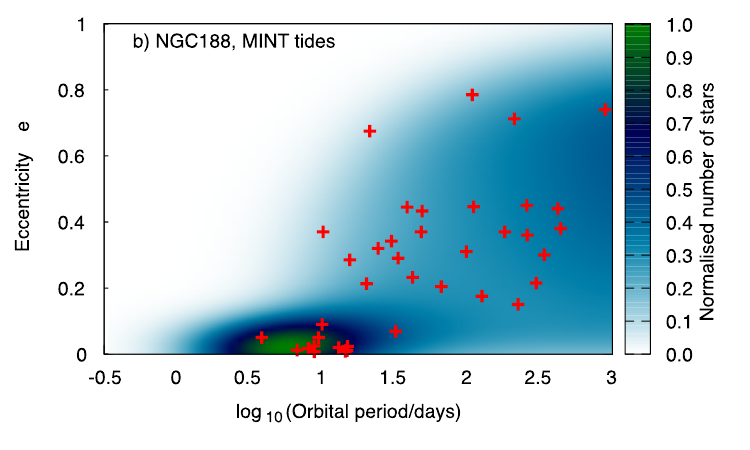}}
    \caption{
    As Fig.~\ref{fig:pleiades} for the NGC 188 cluster.
    }
    \label{fig:ngc188}
\end{figure*}

\end{document}